\documentclass[12pt]{article}

\usepackage{latexsym}
\usepackage{amssymb,amsfonts,amsmath}
\usepackage{graphicx} 
\usepackage{indentfirst}
\usepackage{bbm}
\usepackage{amssymb}
\usepackage{verbatim}
\usepackage{amsmath, amsthm,amssymb}
\usepackage{mathrsfs}
\usepackage{hyperref}
\usepackage{amsfonts}
\usepackage{dsfont}
\usepackage{cite}
\usepackage{tikz}
\usepackage{enumitem}

\parskip=\medskipamount

\newcommand {\cA}{{\cal A}}
\newcommand {\cB}{{\cal B}}

\newcommand {\cD}{{\cal D}}

\newcommand {\cF}{{\cal F}}
\newcommand {\cG}{{\cal G}}

\newcommand {\cI}{{\cal I}}
\newcommand {\cJ}{{\cal J}}
\newcommand {\cK}{{\cal K}}

\newcommand {\cN}{{\cal N}}
\newcommand {\cO}{{\cal O}}

\newcommand {\cQ}{{\cal Q}}

\newcommand {\cZ}{{\cal Z}}

\def\a{\alpha}

\def\b{\beta}

\def\d{\delta}
\def\e{\epsilon}

\def\g{\gamma}

\def\l{\lambda}
\def\m{\mu}

\def\o{\omega}

\def\q{\theta}

\def\s{\sigma}

\def\x{\xi}

\def\F{\Phi}
\def\J{\Psi}
\def\L{\Lambda}

\def\P{\Pi}
\def\Q{\Theta}

\def\ri{{\rm i}}

\def\ra{{\rm a}}

\newcommand{\ad}{{\dot{\alpha}}}                           
\newcommand{\bd}{{\dot{\beta}}}                            
\newcommand{\ve}{\varepsilon}                            

\newcommand{\pa}{\partial}                           
\newcommand{\hf}{\frac12}

\newcommand{\be}{\begin{equation}}
\newcommand{\ee}{\end{equation}}
\newcommand{\bea}{\begin{eqnarray}}
\newcommand{\eea}{\end{eqnarray}}
\newcommand{\non}{\nonumber}

\newcommand{\bm}[1]{\mbox{\boldmath$#1$}}

\def\double #1{#1{\hbox{\kern-2pt $#1$}}}

\newcommand{\gd}{{\dot\g}}

\newif\ifdtup

\def\la{{\langle}}
\def\ra{{\rangle}}
\def\fn3{{{X}_{3}, {\Q}_{3}, \bar {\Q}_{3}}}
\def\fxq{{{X}, {\Q}, \bar {\Q}}}
\def\corr1{{\langle
J_{\a(r_1) \ad(r_1)} (z_1) \, J'{}_{\b(r_2) \bd(r_2)}(z_2)\,  J''{}_{\g(r_3) \gd(r_3)}(z_3)
\rangle}}

\newcommand{\bsubeq}{\begin{subequations}}
\newcommand{\esubeq}{\end{subequations}}

\numberwithin{equation}{section}

\newcommand{\sU}{\mathsf{U}}

\begin{document}

\begin{titlepage}
\begin{flushright}
August, 2022 \\
\end{flushright}
\vspace{5mm}

\begin{center}
{\Large \bf 
Three-point functions of higher-spin supercurrents in 4D ${\cN}=1$ superconformal field theory}
\\ 
\end{center}

\begin{center}

{\bf
Evgeny I. Buchbinder${}^{a}$,
Jessica Hutomo${}^{b}$, and
\\Gabriele Tartaglino-Mazzucchelli${}^{b}$
} \\
\vspace{5mm}

\footnotesize{
${}^{a}$
{\it Department of Physics M013, The University of Western Australia\\
35 Stirling Highway, Crawley W.A. 6009, Australia
}
 \\~\\
${}^{b}$
{\it 
School of Mathematics and Physics, University of Queensland,
\\
 St Lucia, Brisbane, Queensland 4072, Australia}
}
\vspace{2mm}
~\\
\texttt{evgeny.buchbinder@uwa.edu.au, j.hutomo@uq.edu.au, g.tartaglino-mazzucchelli@uq.edu.au}\\
\vspace{2mm}

\end{center}

\begin{abstract}
\baselineskip=14pt

We develop a general formalism to study the three-point correlation functions of conserved higher-spin supercurrent multiplets $J_{\alpha(r) \dot{\alpha}(r)}$ in 4D ${\cal N}=1$ superconformal theory. All the constraints imposed by ${\cal N}=1$ superconformal symmetry on the three-point function $\langle J_{\alpha(r_1) \dot{\alpha}(r_1)} J_{\beta(r_2) \dot{\beta}(r_2) }J_{\gamma(r_3) \dot{\gamma}(r_3)}\rangle$ are systematically derived for arbitrary $r_1, r_2, r_3$, thus reducing the problem mostly to computational and combinatorial. As an illustrative example, we explicitly work out the allowed tensor structures contained in $\langle J_{\alpha(r) \dot{\alpha}(r)} J_{\beta \dot{\beta} } J_{\gamma \dot{\gamma}}\rangle$, where $J_{\alpha \dot{\alpha}}$ is the supercurrent. We find that this three-point function depends on two independent tensor structures, though the precise form of the correlator depends on whether $r$ is even or odd. The case $r=1$ reproduces the three-point function of the ordinary supercurrent derived by Osborn. Additionally, we present the most general structure of mixed correlators of the form $\langle L L J_{\alpha(r) \dot{\alpha}(r)}\rangle$ and $\langle J_{\alpha(r_1) \dot{\alpha}(r_1)} J_{\beta(r_2) \dot{\beta}(r_2)} L \rangle$, where $L$ is
the flavour current multiplet.
\end{abstract}
\vspace{5mm}

\vfill
\end{titlepage}

\newpage
\renewcommand{\thefootnote}{\arabic{footnote}}
\setcounter{footnote}{0}

\tableofcontents{}
\vspace{1cm}
\bigskip\hrule

\allowdisplaybreaks


\section{Introduction}


Finding how conformal symmetry constrains correlation functions of local, primary operators is one of the basic and most important 
questions in conformal field theory (see refs.~\cite{Polyakov:1970xd, Schreier:1971um, Migdal:1971xh, Migdal:1972tk, Ferrara:1972cq, Ferrara:1973yt, Koller:1974ut, Mack:1976pa, Stanev:1988ft, Fradkin:1978pp} for early results). Among primary fields (corresponding to lowest weight states of the conformal 
algebra) there is an important class of operators which are conserved currents. They correspond to lowest weight states whose scaling dimension saturates 
the unitarity bound.
The systematic approach to study correlation functions of conserved currents was developed in~\cite{OP, EO}\footnote{The analyses of~\cite{OP, EO} 
were performed in general dimensions and did not consider parity-violating structures relevant for three-dimensional conformal field theories. 
These structures were later found in~\cite{Giombi:2011rz}.} and later extended to superconformal field theories in various dimensions \cite{Park1, OsbornN1, Park, Park6, Park3, KT, Nizami:2013tpa, Buchbinder:2015qsa, 
Buchbinder:2015wia, Kuzenko:2016cmf, Buchbinder:2021gwu, Buchbinder:2021izb, Buchbinder:2021kjk, Buchbinder:2021qlb, Jain:2022izp}.
Several novel approaches to the construction of correlation functions of conserved currents which carry out the calculations in momentum space, 
using methods such as spinor-helicity variables were studied in~\cite{Jain:2020puw, Jain:2020rmw, Jain:2021gwa, Jain:2021vrv, Jain:2021wyn, Isono:2019ihz, Bautista:2019qxj}.
The most studied examples of conserved currents in (super)conformal field theory are the energy-momentum tensor and vector currents; their most general 
three-point functions were determined in~\cite{OP, EO}. However, (super)conformal field theories can possess higher-spin conserved (super)currents. 
The general structure of three-point functions of conserved higher-spin, bosonic, vector currents was obtained by Stanev~\cite{Stanev:2012nq} and 
Zhiboedov~\cite{Zhiboedov:2012bm}, see~\cite{Elkhidir:2014woa} for similar results using the embedding 
formalisms~\cite{Weinberg:2010fx, Weinberg:2012mz, Costa:2011dw, Costa:2011mg, Costa:2014rya, Fortin:2020des} (and \cite{Goldberger:2011yp, Goldberger:2012xb, Fitzpatrick:2014oza,Khandker:2014mpa} for supersymmetric extensions).\footnote{Correlation functions of fermionic operators were discussed in~\cite{Elkhidir:2014woa, BS22}.} Recently, Ref.~\cite{KT-M21} proposed supertwistor realisations of $(p,q)$ anti-de Sitter (AdS) superspaces in three dimensions and $\cN$-extended AdS superspaces in four dimensions, see \cite{Binder:2020raz} for the non-supersymmetric case. These formulations might be useful for studying correlation functions on AdS.

As was proven by Maldacena and Zhiboedov in~\cite{Maldacena:2011jn} (see also~\cite{Stanev:2013qra, Alba:2013yda, Alba:2015upa}),
all correlation functions of higher-spin conserved currents are equal to those of a free theory. The main assumption of the Maldacena--Zhiboedov theorem was that the conformal field theory under consideration possesses a unique conserved current of spin 2, 
the energy-momentum tensor. However, in~\cite{Maldacena:2011jn} it was also shown that if a conformal field theory possesses a conserved 
fermionic higher-spin current then it has an additional conserved current of spin 2. This implies that the Maldacena--Zhiboedov theorem does not seem 
to hold in superconformal field theories possessing higher-spin currents.

In supersymmetric theories, the bosonic and fermionic currents form supermultiplets. For example, the energy-momentum tensor
together with the fermionic supersymmetry current belong to the supercurrent 
multiplet \cite{FZ} (see also \cite{GGRS, MSW, KS2, K-var, K-var1}). On the other hand, an additional conserved current of spin 2 will belong 
to the supermultiplet of superspin 2 which also includes a higher-spin current of spin $\frac{5}{2}$, see~\cite{Buchbinder:2021qlb} for a 
similar discussion in three dimensions. In general, a superconformal field theory possessing bosonic higher-spin currents will also possess fermionic ones.  
As discussed above, this naturally leads to a violation of the main assumption of the Maldacena-Zhiboedov theorem. 
Thus, correlation functions of higher-spin conserved currents in superconformal field theories do not have to 
coincide with those in free theory which makes their study particularly interesting. 

The study of the most general form of three-point correlation functions of higher-spin conserved currents in four-dimensional (4D) $\cN =1$ superconformal field theory was initiated
in~\cite{Buchbinder:2021kjk}, 
where the three-point functions of higher-spin \textit{spinor} currents $S_{\alpha(k)}$ were obtained.\footnote{We use the standard notation  
$S_{\alpha(k)}= S_{(\alpha_1 \dots \alpha_k)}$.} In this paper, we continue our study to explore how $\cN = 1$ superconformal symmetry and conservation laws
constrain the structure of three-point functions involving the conformal higher-spin current multiplet $J_{a_1 \dots a_r} = J_{(a_1 \dots a_r)} = J_{a(r)}$. This real symmetric and traceless rank-$r$ tensor superfield $J_{a(r)}$ 
is in one-to-one correspondence
with the real\footnote{Since $J_{\a(r) \ad(r)}$ is real, it is neutral under the $R$-symmetry group $\sU(1)_{R}$.} totally symmetric spinor superfield $J_{\a_1 \dots \a_r \ad_1 \dots \ad_r} = J_{(\a_1 \dots \a_r)(\ad_1 \dots \ad_r)} \equiv J_{\a(r) \ad(r)}$ via
\bea
J_{\a(r) \ad(r)} = (\s^{a_1})_{\a_1 \ad_1} \dots (\s^{a_r})_{\a_r \ad_r} J_{a_1 \dots a_r}~.
\label{z1}
\eea
We will refer to \eqref{z1} as higher-spin supercurrent. The superfield $J_{\a(r) \ad(r)}$, which belongs to the representation $(r/2, r/2)$ of the Lorentz group, is primary with conformal weight $(q, \bar{q}) = (\frac{r+2}{2}, \frac{r+2}{2})$ and scaling dimension $r+2$. It is subject to the conservation equations
\bea
D^{\b} J_{\b \a(r-1) \ad(r)} = 0~, \qquad 
\bar{D}^{\bd} J_{\a(r) \bd \ad(r-1)}=0~,
\label{z2}
\eea
where $D^{\b}, \bar{D}^{\bd}$ are the superspace covariant derivatives. 
The case $r=1$ corresponds to the ordinary conformal supercurrent \cite{FZ}. The case $r>1$ was first described in \cite{HST}.

The structure of the higher-spin current multiplets, including non-conformal ones, have been a subject of several works. For instance, explicit realisations in terms of free scalar and vector multiplets are known for $J_{\a(r) \ad(r)}$, in Minkowski \cite{KMT, Buchbinder:2017nuc, Hutomo:2017phh, Hutomo:2017nce, Koutrolikos:2017qkx, Buchbinder:2018wwg, Buchbinder:2018gle}, AdS \cite{BHK18} and conformally-flat backgrounds \cite{Kuzenko:2022hdv}. Higher-spin supercurrent $J_{\a(r) \ad(r)}$ may also be realised in terms of the on-shell, gauge-invariant, chiral field strengths $W_{\a(r)} $ obeying
$
\bar D_\bd W_{\a(r)} =0$ and $D^\b W_{\b \a(r-1)} =0$~. Its realisation is given by \cite{BHK18, BGK18, GK19}:
\bea
J_{\a(r) \ad(r) } =W_{\a(r)} \bar W_{\ad(r)} ~.
\label{zz-b}
\eea
Here $W_\a$, $W_{\a(2)}$ and $W_{\a(3)}$ are the gauge-invariant field strengths describing the on-shell vector, gravitino and linearised supergravity multiplets, respectively.  For $r>3$, the superfields $W_{\a (r)}$ are the on-shell gauge-invariant field strengths corresponding to the massless higher-spin gauge multiplets 
\cite{KPS,KS93} (see section 6.9 of \cite{Ideas} for a review). Setting $r=1$ in 
\eqref{zz-b} gives the supercurrent of the free $\cN=1$ vector multiplet, 
$J_{\a \ad } =W_{\a} \bar W_{\ad}$.

The main aim of  this work is to develop a general formalism to study the three-point correlation function of conserved higher-spin 
supercurrents 
\be
\la J_{\a(r_1) \ad(r_1)} (z_1) J_{\b(r_2) \bd(r_2) }(z_2) J_{\g(r_3) \gd(r_3)} (z_3)\ra
\label{z3}
\ee
in 4D $\cN=1$ superconformal field theory.
We use a hybrid approach which combines 
the group theoretic formalism of Osborn~\cite{OsbornN1} with the index-free method based on using auxiliary vectors/spinors. The latter enables us to encode symmetric traceless tensors by polynomials in commuting auxiliary spinors. The approach based on auxiliary polarisation vectors is widely used in the literature, e.g. \cite{Giombi:2011rz, Costa:2011mg, Stanev:2012nq, Zhiboedov:2012bm, Nizami:2013tpa, Elkhidir:2014woa} (for earlier work see also e.g. Ref. \cite{Craigie:1983fb} and references therein), where one usually has to work with the spacetime points explicitly when analysing conservation laws. In contrast, by making use of the general formalism of Osborn \cite{OsbornN1}, all information about the three-point function \eqref{z3} is concentrated in
two polynomials, both of which are functions of a \textit{single} superconformally covariant \textit{bosonic} vector $X$, and the auxiliary commuting spinors. 
Within our approach, the constraints imposed by conservation laws and symmetries with respect to  permutations of superspace points take a simple form; hence, they can be efficiently implemented and solved computationally in \textit{Mathematica}.
Our method is a supersymmetric extension of the generating function formalism recently proposed in \cite{BS22}. Once the formalism is developed, finding the general solution of the correlation function~\eqref{z3} 
to a large extent becomes a computational and combinatorial (though still quite a difficult) problem. One of the complications involves determining various linear dependence relations in the possible set of solutions. This problem will be discussed elsewhere.
We believe that our formalism can be generalised to the case of the most general higher-spin conformal current multiplets containing both vector and spinor indices, which are of the form $J_{\a(m) \ad(n)}$ corresponding to Lorentz type
$(m/2, n/2)$, with $m, n > 1$ and $m \neq n$ \cite{KR}. This problem will also be studied elsewhere. 

The paper is organised as follows. In section \ref{Section2} we review the general construction of two- and three-point functions of conserved currents mostly following~\cite{OsbornN1}. 
In section \ref{section3} we introduce the formalism aimed at computing the three-point function~\eqref{z3} for arbitrary $r_1, r_2, r_3$. In the remaining sections, we demonstrate the efficiency of our formalism by computing several new mixed correlation functions of $J_{\a(r) \ad(r)}$ with the supercurrent and the flavour current multiplet $L$. The latter is a 
real scalar superfield containing the vector current $V_m$. 
More precisely, in section \ref{section4}, we consider 
$\la J_{\a(r) \ad(r)} (z_1) \, J_{\b \bd}(z_2)\,  J_{\g \gd}(z_3) \ra$ and find the most 
general solutions in an explicit form for all values of $r$. In particular, for $r=1$ we reproduce the three-point function of the supercurrent found previously in~\cite{OsbornN1}. For higher $r$, we construct a generating function which produces all possible linearly dependent solutions. The number of linearly independent tensor structures compatible with all the constraints are then found computationally. We find that the correlator depends on two independent tensor structures, though its precise form depends on the values of $r$ (even or odd). 
In section \ref{section5}, we generalise our formalism to the mixed three-point functions involving $J_{\a(r) \ad(r)}$ and the flavour current multiplet. 
We find the general solutions for $\la L(z_1) L(z_2) J_{\a(r) \ad(r)} (z_3)\ra$ and $\la J_{\a(r_1) \ad(r_1)}(z_1) J_{\b(r_2) \bd(r_2)}(z_2) L(z_3)\ra$, for all values of $r, r_1, r_2$ in an explicit form. In the former case, the correlator $\la LLJ \ra$ depends on a single tensor structure. In the $\la JJL \ra$ case, the correlator depends on two tensor structures if $r_1 < r_2$; and a single tensor structure if $r_1 = r_2$.


\section{Superconformal building blocks} \label{Section2}


This section contains a concise summary of two- and three-point superconformal building blocks in four-dimensional $\cN=1$ superspace, which are essential for our analysis. These superconformal structures were introduced in \cite{Park1, OsbornN1}, and later generalised to arbitrary ${\cN}$ in \cite{Park} (see also \cite{KT} for a review). 
Our notation and conventions are those of \cite{Ideas}.


\subsection{Two-point structures}


We denote the $\cN=1$ Minkowski superspace by ${\mathbb M}^{4|4}$. It is parametrised by coordinates $z^A = (x^a, \q^\a, {\bar \q}_\ad)$, where $a = 0,1, 2, 3;~ \a, \ad = 1, 2 $. Let $z_1$ and $z_2$ be two different points in superspace. 
All building blocks for the two- and three-point correlation functions are composed of the two-point structures:
\bsubeq
\bea
x^a_{\bar{1} 2} &=& -x^{a}_{2 \bar{1}} = x^{a}_{1-} - x^a_{2+} + 2 \ri \,\q_{2 }\s^{a} {\bar \q}_{1}~,\\
\q_{12} &=& \q_1 - \q_2~, \qquad {\bar \q}_{12} = {\bar \q}_1 - {\bar \q}_2~,
\eea 
\esubeq
which are invariants of the $Q$-supersymmetry transformations. Here $x^a_{\pm}= x^a \pm \ri \theta \sigma^a {\bar \theta}$.
In spinor notation, we write
\bsubeq 
\bea
{x}_{\bar{1} 2}{}^{\ad \a} &=& (\tilde{\s_a})^{\ad \a}x^{a}_{\bar{1} 2}~, \\
{x}_{2 \bar{1} \,\a \ad} &=& (\s_a)_{\a \ad} \, x^a_{2 \bar{1} }=
-(\s_a)_{\a \ad} \, x^a_{\bar{1} 2}= - \ve_{\a \b} \ve_{\ad \bd} {x}_{\bar{1} 2}{}^{\bd \b}~, \label{eps-x}\\
{x}_{\bar{1} 2}^{\,\dagger}{}^{\ad \a} &=& - {x}_{\bar{2} 1}{}^{\ad \a}~. \label{hc-x}
\eea
\esubeq
Note that ${x}_{\bar{1} 2}{}^{\ad \a} x_{2 \bar{1} \, \a \bd} = x_{\bar{1} 2}{}^{2}\, \d^{\ad}{}_{\bd}$. 
We sometimes employ matrix-like conventions of \cite{OsbornN1, KT} where the spinor indices are not explicitly written:
\bsubeq
\bea
&&
\psi = (\psi^{\a})~, \quad  \tilde{\psi} = (\psi_{\a})~, \quad  \bar{\psi} = (\bar{\psi}^{\ad})~, \quad \tilde{\bar{\psi}} = (\bar{\psi}_{\ad})~, 
\label{e2a}
\\
&&
x = (x_{\a \ad})~, \quad \tilde{x} = (x^{\ad \a})~.
\label{e2b}
\eea
\esubeq
Since $x^2 \equiv x^a x_a = -\hf {\rm tr}(\tilde{x} x)$, it follows that $\tilde{x}^{-1} = -x/x^2$.
The notation `$\tilde{x}_{\bar{1} 2}$' means that $\tilde{x}_{\bar{1} 2}$ is antichiral with respect to $z_1$ and chiral with respect to $z_2$. 
That is,
\be 
D_{(1)  \a} \tilde{x}_{\bar{1} 2}=0\,, \qquad \bar D_{(2) \ad} \tilde{x}_{\bar{1} 2}=0~,
\label{E1}
\ee
where $D_{(1) \a}$ and $\bar D_{(1) \ad}$  are the superspace covariant spinor derivatives acting on the point $z_1$. Similarly,
$D_{(2) \a}$ and $\bar D_{(2) \ad}$ act on the point $z_2$. Explicitly, it holds that 
\bea
D_{\a} = \frac{\pa}{\pa \q^{\a}}+ \ri (\s^a)_{\a \ad} \bar{\q}^{\ad} \frac{\pa}{\pa x^a}~, \qquad \bar D_{\ad} = -\frac{\pa}{\pa \bar \q^{\ad}} - \ri {\q}^{\a} (\s^a)_{\a \ad}  \frac{\pa}{\pa x^a}~.
\eea
The $Q$-supersymmetry generators are defined as
\bea
Q_{\a} =  \ri \frac{\pa}{\pa \q^{\a}}+(\s^a)_{\a \ad} \bar{\q}^{\ad} \frac{\pa}{\pa x^a}~, \qquad 
\bar Q_{\ad} = -\ri \frac{\pa}{\pa \bar \q^{\ad}} - {\q}^{\a} (\s^a)_{\a \ad}  \frac{\pa}{\pa x^a}~,
\eea
and thus we have that
\bea
\{ D_{\a}, Q_{\b} \} = \{ D_{\a}, \bar Q_{\bd} \} = \{ \bar{D}_{\ad}, Q_{\b} \}= \{ \bar D_{\ad}, \bar Q_{\bd} \} = 0~.
\eea
Indeed, it can be checked that $\tilde{x}_{\bar{1} 2}$ is annihilated by the supercharge operators,
\bea
Q_{(1)\a} \tilde{x}_{\bar{1} 2} = \bar{Q}_{(1)\ad} \tilde{x}_{\bar{1} 2} = Q_{(2)\a} \tilde{x}_{\bar{1} 2} = \bar{Q}_{(2)\ad} \tilde{x}_{\bar{1} 2} = 0~.
\eea

Let us define the normalised two-point functions \cite{OsbornN1},
\bea
\hat{x}_{2 \bar{1} \,\a \ad} = \frac{x_{2 \bar{1} \a \ad}}{(x_{\bar{1} 2}{}^2)^{1/2}}~, \qquad  
\hat{x}_{\bar{1} 2}{}^{\ad \a} = \frac{{x}_{\bar{1} 2}{}^{\ad \a}}{(x_{\bar{1} 2}{}^2)^{1/2}}~.
\label{I-def}
\eea 
They satisfy
\bea
\hat{x}_{\bar{1} 2}{}^{\ad \a} 
\hat{x}_{2 \bar{1} \,\b \ad} = \d^{\a}{}_{\b}~, \qquad 
\hat{x}_{\bar{1} 2}{}^{\ad \a} \hat{x}_{2 \bar{1} \,\a \bd} = \d^{\ad}{}_{\bd}~.
\label{I-inv}
\eea
In accordance with \cite{OsbornN1}, one can construct the vector representation of the inversion tensor in terms of the spinor two-point functions \eqref{I-def} as follows:
\bsubeq \label{vect-2pt}
\bea
&&I_{ab} (x_{1 \bar 2}, x_{\bar{1} 2}) = \bar{I}_{ba}(x_{\bar{2} 1}, x_{2 \bar{1}})=
\hf {\rm tr}\,(\tilde{\s}_a \,\hat{x}_{1 \bar{2}} \,\tilde{\s}_b\, \hat{x}_{2 \bar{1}})~,\\
&&I_{ac} (x_{1 \bar 2}, x_{\bar{1} 2}) \bar{I}^{cb}(x_{\bar{2} 1}, x_{2 \bar{1}}) = \d_{a}{}^{b}~.
\eea
\esubeq
In the purely bosonic case, $I_{ab} (x_{1 \bar 2}, x_{\bar{1} 2})$ reduces to the conformal inversion tensor \cite{OP}
\bea
I_{ab} (x_{12})= \eta_{a b} - \frac{2}{x_{12}^2} x_{12 a}{} x_{12 b}~,
\eea
which played a pivotal role in studying conformal invariance in arbitrary dimensions \cite{OP, EO}.

We can also construct higher-spin extensions of the above operators, which act on the space of symmetric traceless tensors of arbitrary rank. Specifically, we define
\bea
\cI_{\a(k) \ad(k)} (x_{2 \bar{1}}) &:=& 
\hat{x}_{2 \bar{1} (\a_1 (\ad_1} \dots \hat{x}_{2 \bar{1} \,\a_k) \ad_k)}~,
\label{I-hs}
\eea
along with its inverse
\bea
\bar \cI^{\ad{(k)} \a(k)} (\tilde{x}_{\bar{1} 2}) := 
\hat{x}_{\bar{1} 2}{}^{(\ad_1 (\a_1} \dots \hat{x}_{\bar{1} 2}{}^{\ad_k) \a_k)}~.
\label{I-hs-up}
\eea
Due to the properties \eqref{eps-x} and \eqref{hc-x}, it holds that
\bsubeq
\bea
\cI_{\a(k) \ad(k)} (x_{2 \bar{1}}) &=&
(-1)^k \ve_{\a_1 \b_1} \dots \ve_{\a_k \b_k} \ve_{\ad_1 \bd_1} \dots \ve_{\ad_k \bd_k} \bar \cI^{\bd{(k)} \b(k)} (\tilde{x}_{\bar{1} 2})~,\\
\Big(\bar \cI^{\ad{(k)} \a(k)} (\tilde{x}_{\bar{1} 2}) \Big)^{\dagger} &=& 
(-1)^k \,\bar \cI^{\ad{(k)} \a(k)} (\tilde{x}_{\bar{2} 1})~.
\eea
\esubeq
As a generalisation of \eqref{I-inv}, we have that
\bsubeq
\bea
&&\cI_{\a(k) \ad(k)}(x_{2 \bar{1}}) \bar \cI^{\ad{(k)} \g(k)} (\tilde{x}_{\bar{1} 2}) = \d_{(\a_1}^{(\g_1} \dots \d_{\a_k)}^{\g_k)}~,\\
&&\bar \cI^{\gd{(k)} \g(k)} (\tilde{x}_{\bar{1} 2})\cI_{\g(k) \ad(k)}(x_{2 \bar{1}})  = \d_{(\ad_1}^{(\gd_1} \dots \d_{\ad_k)}^{\gd_k)}~.
\eea
\esubeq
Since all superconformal transformations may be generated by combining inversions with
ordinary supersymmetry, the operators \eqref{I-hs} and \eqref{I-hs-up} play a crucial role in constructing correlation functions for primary operators with arbitrary spin.

We also note several useful differential identities:
\bsubeq \label{2ptids}
\bea
&&D_{(1) \a} (x_{\bar{2} 1})^{\bd \b} = 4 \ri \d_{\a}{}^{\b} \bar{\q}_{12}^{\bd}~, \qquad \bar{D}_{(1) \ad}\, (x_{\bar{1} 2})^{\bd \b} = 4 \ri \d_{\ad}{}^{\bd} \q_{12}^{\b}~, \\
&&D_{(1) \a} \bigg(\frac{1}{x_{\bar{2} 1}{}^{2}} \bigg) = - \frac{4 \ri}{x_{\bar{2} 1}{}^2}  (\tilde{x}_{\bar{2} 1}{}^{-1})_{\a \bd} \bar{\q}_{12}^{\bd}~, \quad \bar D_{(1) \ad} \bigg(\frac{1}{x_{\bar{1} 2}{}^{2}} \bigg) = - \frac{4 \ri}{x_{\bar{1} 2}{}^2}  (\tilde{x}_{\bar{1} 2}{}^{-1})_{\a \ad} {\q}_{12}^{\a}~.~~~~~
\eea
\esubeq
Here and throughout, we assume that the superspace points are not coincident, $z_1 \neq z_2$. 

Consider a tensor superfield $\cO^{\cA}(z)$ transforming in a representation $T$ of the Lorentz group with respect to the index $\cA$.\footnote{We assume the representations $T$ is irreducible. The superscript `$\cA$' collectively denotes the undotted and dotted spinor indices on which the Lorentz generators act.} Such a superfield is called primary if its infinitesimal superconformal transformation law reads
\be
\begin{aligned}
\d\, \cO^\cA(z) &= - \x \, \cO^\cA (z) 
+ (\hat{\o}^{\a \b} (z) M_{\a \b}+ 
\hat{ \bar{\o}}^{\dot{\a} \dot{\b}} (z) 
\bar{M}_{\dot{\a} \dot{\b}} )^\cA{}_\cB\,
\cO^\cB (z) \\
& - 2\left( q\, \s(z) + \bar{q}\, \bar{\s} (z) \right) 
\cO^\cA (z)~.
\end{aligned}
\ee
In the above, $\xi$ is the superconformal Killing vector,
\bea
\xi = \bar{\xi} = \xi^a (z) \pa_a + \xi^{\a} (z) D_{\a}+ \bar{\xi}_{\ad} (z) \bar{D}^{\ad}~.
\eea
The superfield parameters $\hat{\o}^{\a \b} (z), ~\s(z)$ correspond to the `local' Lorentz and scale transformations: they are expressed in terms of $\xi^{A} = (\xi^a, \xi^{\a}, \bar \xi_{\ad})$, see \cite{OsbornN1} for details. The weights $q$ and $\bar q$ are such that $(q+\bar q)$ is the scale dimension and $(q-\bar q)$ is proportional to the $\rm U(1)_R$ charge of the superfield $\cO^{\cA}$.

Following the general formalism of \cite{OsbornN1,Park1, Park},
the two-point function of a primary superfield
$\cO^{\cA}$ with its
conjugate $\bar{\cO}^{\cB}$ is given by
\be
\langle \cO^{\cA} (z_1)\;\bar{\cO}^{\cB} (z_2)\rangle
~=~ C_{\cO}\;\frac{ 
\cI^{\cA \cB} ({{x}_{1 \bar{2}}}, {{x}_{2 \bar{1}}}) }
{ (x_{\bar{1}2}{}^2)^{\bar q} (x_{\bar{2}1}{}^2)^q }~,
\label{2pt-gen}
\ee
where $ C_{\cO}$ is an overall normalisation constant and $\cI$ is an appropriate representation of the inversion tensor. 

In this paper, we are interested in the conserved higher-spin supercurrent multiplet $J_{\a(r) \ad(r)}$. Its two-point function is fixed by the superconformal symmetry to the form:
\bea
\langle J_{\a(r) \ad(r)} (z_1)\;J_{\b(r) \bd(r)} (z_2)\rangle
= C_{J} \frac{\cI_{\a(r) \bd(r)} (x_{1 \bar{2}})\, \cI_{\b(r) \ad(r)} (x_{2 \bar{1}})}{(x_{\bar{1} 2}{}^2 \,x_{\bar{2} 1}{}^2)^\frac{r+2}{2}}~,
\eea
where we have use the fact that for $J_{\a(r) \ad(r)}$ we have $q=\bar q= \frac{r+2}{2}$. 


\subsection{Three-point structures}


Given three superspace points $z_1, z_2$ and $z_3$, we have the following three-point structures  ${Z}_1, {Z}_2$ and ${Z}_3$, with ${Z}_1 = ({X}_1^a, {\Q}_1^{\a}, {\bar \Q}_{1}^{\ad})$ (see \cite{Park1, OsbornN1} for details):
\bsubeq
\bea
&&{X}_{1} = \tilde{x}^{-1}_{\bar{2}1} \tilde{x}_{\bar{2} 3} \tilde{x}^{-1}_{\bar{1} 3}~, \,\,\, \tilde{{\Q}}_{1} = \ri \big(\tilde{x}^{-1}_{\bar{2} 1} \bar{\q}_{12} -  \tilde{x}^{-1}_{\bar{3} 1} \bar{\q}_{13}\big)~, \,\,\, \tilde{\bar{{\Q}}}_{1} = \ri \big({\q}_{12} \tilde{x}^{-1}_{\bar{1} 2} \ - {\q}_{13} \tilde{x}^{-1}_{\bar{1} 3} \big)~,~~~~~ \label{Z1}\\
&&{X}_{2} = \tilde{x}^{-1}_{\bar{3}2} \tilde{x}_{\bar{3} 1} \tilde{x}^{-1}_{\bar{2} 1}~, \,\,\, \tilde{{\Q}}_{2} = \ri \big(\tilde{x}^{-1}_{\bar{3} 2} \bar{\q}_{23} -  \tilde{x}^{-1}_{\bar{1} 2} \bar{\q}_{21}\big)~, \,\,\, \tilde{\bar{{\Q}}}_{2} = \ri \big({\q}_{23} \tilde{x}^{-1}_{\bar{2} 3} \ - {\q}_{21} \tilde{x}^{-1}_{\bar{2} 1} \big)~,~~~~~ \label{Z2}\\
&&{X}_{3} = \tilde{x}^{-1}_{\bar{1}3} \tilde{x}_{\bar{1} 2} \tilde{x}^{-1}_{\bar{3} 2}~, \,\,\, \tilde{{\Q}}_{3} = \ri \big(\tilde{x}^{-1}_{\bar{1} 3} \bar{\q}_{31} -  \tilde{x}^{-1}_{\bar{2} 3} \bar{\q}_{32}\big)~, \,\,\, \tilde{\bar{{\Q}}}_{3} = \ri \big({\q}_{31} \tilde{x}^{-1}_{\bar{3} 1} \ - {\q}_{32} \tilde{x}^{-1}_{\bar{3} 2} \big)~.~~~~~ \label{Z3}
\eea
\esubeq
Since \eqref{Z2} and \eqref{Z3} are obtained through cyclic permutations of superspace points, it suffices to study the properties of \eqref{Z1}. Let us also define 
\bea
\bar{{X}}_{1} = {X}_1^{\dagger} &=& -\tilde{x}_{\bar{3}1}{}^{-1} \tilde{x}_{\bar{3} 2} \tilde{x}_{\bar{1} 2}{}^{-1}~.
\eea
Similar relations hold for $\bar{{X}}_{2}, \bar{{X}}_{3}$.

We list several properties of ${Z}$'s which will be useful later
\bsubeq \label{XbarX}
\bea
&&{X}_1^2 = \frac{x_{\bar{2} 3}{}^2}{x_{\bar{2} 1}{}^2 x_{\bar{1} 3}{}^2}~, \qquad {\bar X}_1^2 = \frac{x_{\bar{3} 2}{}^2}{x_{\bar{3} 1}{}^2 x_{\bar{1} 2}{}^2}~, \\
&&{\bar X}_{\a \ad} = {X}_{\a \ad} + \ri {P}_{\a \ad}~, \quad {P}_{\a \ad} = -4 {\Q}_{\a} {\bar \Q}_{\ad}~, \quad P_a P_b = \frac{1}{4} \eta_{a b}{P}^2 = -2 \eta_{ab} {\Q}^2 {\bar \Q}^2~,~~~~~~ \label{Xbar-def}\\
&&\frac{1}{{\bar X}^{2k}} = 
\frac{1}{{X}^{2k}} -2\ri k \frac{({P} \cdot {X})}{{X}^{2k+2}} - \frac{k}{2} (k-1) \frac{P^2}{X^{2k+2}}~, 
\quad ({P} \cdot {X}) = -\hf {P}^{\ad \a} {X}_{\a \ad}~,
\eea
\esubeq
where, throughout the paper, we adopt the notation $X^k \equiv (X^2)^{k/2}$.

In particular, we see that ${\bar X}$ is not an independent variable for it can be expressed in terms of 
${X}, {\Q}, {\bar \Q}$.
Variables ${Z}$ with different labels are related to each other via the identities
\bsubeq \label{Z13}
\bea
\tilde{x}_{\bar{1} 3} {X}_3 \tilde{x}_{\bar{3} 1} &=& - \bar{{X}}_1^{-1} = \frac{\tilde {\bar{ {X}}}_1}{\bar {{X}}_1^2}~, \qquad \tilde{x}_{\bar{1} 3} \bar{{X}}_3 \tilde{x}_{\bar{3} 1} = - {X}_1^{-1} = \frac{\tilde { {X}}_1}{{X}_1^2}~, \\
\frac{x_{\bar{3} 1}{}^2}{x_{\bar{1} 3}{}^2} \tilde{x}_{\bar{1} 3} \tilde{{\Q}}_3 &=&
 - {X}_1^{-1} \tilde{{\Q}}_1~, \qquad \frac{x_{\bar{1} 3}{}^2}{x_{\bar{3} 1}{}^2}  \tilde{\bar {{\Q}}}_3 \tilde{x}_{\bar{3} 1} =
\tilde{\bar{{\Q}}}_1 \bar{{X}}_1^{-1} ~.
\eea
\esubeq
It is also convenient to define the normalised three-point building block $\hat{X}_{\a \ad}$
\bea
\hat{X}_{\a \ad} = \frac{{X}_{\a \ad}}{({X}^2)^{1/2}}~.
\eea
We then construct the higher-spin operator
\bea
\cI_{\a(k) \ad(k)} (X) = \hat{X}_{(\a_1 (\ad_1} \dots \hat{X}_{\a_k) \ad_k)}~,
\eea
along with its inverse
\bea
\bar{\cI}^{\ad(k) \a(k)}(X) = \hat{X}^{(\a_1 (\ad_1} \dots \hat{X}^{\a_k) \ad_k)}~,
\eea
with properties
\bsubeq
\bea
&&\cI_{\a(k) \ad(k)}(X) \bar \cI^{\ad{(k)} \g(k)} (X) = (-1)^k\d_{(\a_1}^{(\g_1} \dots \d_{\a_k)}^{\g_k)}~,\\
&&\bar \cI^{\gd{(k)} \g(k)} (X)\cI_{\g(k) \ad(k)}(X)  = (-1)^k \d_{(\ad_1}^{(\gd_1} \dots \d_{\ad_k)}^{\gd_k)}~.
\eea
\esubeq
The normalised, higher-spin operator for $\bar{X}$ can also be defined in a similar way:
\bea
\cI_{\a(k) \ad(k)} (\bar{X}) = \hat{\bar{X}}_{(\a_1 (\ad_1} \dots \hat{\bar X}_{\a_k) \ad_k)}~.
\eea
In the vector representation, we can also define
\bea
I_{ab}(\bar{X}, X) = -\hf {\rm tr} \big( \tilde{\s}_{a}\, \hat{\bar{X}} \,\tilde{\s}_{b} \,\hat{X} \big)~.
\eea

Various primary superfields, including conserved current multiplets, are subject to certain differential constraints. These need to be taken into account when constraining correlation functions. 
For three-point functions, the action of covariant spinor derivatives on an arbitrary function $t( X_3, \Q_3, \bar{\Q}_3)$ can be simplified using these useful differential identities \cite{OsbornN1}:
\bsubeq \label{Cderivs}
\bea
D_{(1)\a} \,t(\fn3) &=& -\frac{\ri}{x_{\bar{1} 3}{}^2}(x_{1 \bar{3}})_{\a \ad} \bar{\cD}_{(3)}^{\ad} t(\fn3)~,\\
\bar{D}_{(1)\ad} \,t(\fn3) &=& -\frac{\ri}{x_{\bar{3} 1}{}^2}(x_{3 \bar{1}})_{\a \ad} {\cD}_{(3)}^{\a} t(\fn3)~,\\
D_{(2)\a}\, t(\fn3) &=& \frac{\ri}{x_{\bar{2} 3}{}^2}(x_{2 \bar{3}})_{\a \ad} \bar{\cQ}_{(3)}^{\ad} t(\fn3)~, \label{Cderiv-2a}\\
\bar{D}_{(2) \ad}\, t(\fn3) &=& \frac{\ri}{x_{\bar{3} 2}{}^2}(x_{3 \bar{2}})_{\a \ad} {\cQ}_{(3)}^{\a} t(\fn3)~, \label{Cderiv-2b}
\eea
\esubeq
where, for $( X_3, \Q_3, \bar{\Q}_3) \longrightarrow (\fxq)$;~$\cD_{(3)}, \bar{\cD}_{(3)} \longrightarrow \cD, \bar{\cD}$ and $\cQ_{(3)}, \bar{\cQ}_{(3)}\longrightarrow \cQ, \bar{\cQ}$, we define the conformally covariant 
operators \cite{OsbornN1}
\be
\begin{aligned}
&\cD_{\bar A} = (\pa / \pa X^a, \cD_{\a}, \bar{\cD}^{\ad})~, \qquad \cQ_{\bar A} = (\pa / \pa X^a, {\cal Q}_{\a},
\bar{\cal Q}^{\ad})~, \\
&\cD_{\a} = \frac{\pa}{ \pa {\Q}^{\a} }
-2{\rm i}\,(\s^a)_{\a \ad} \bar{\Q}^\ad
\frac{\pa }{ \pa X^a }~, \qquad
\bar{\cD}^{\ad} = \frac{\pa}{ \pa \bar{\Q}_{\ad} }~, \\
&{\cal Q}_{\a}  =  \frac{\pa}{ \pa {\Q}^{\a} }~, \qquad
\bar{\cal Q}^{\ad} = \frac{\pa}{ \pa \bar{\Q}_{\ad} } 
+ 2{\rm i}\, {\Q}_{\a} (\tilde{\s}^a)^{\ad \a}
\frac{\pa}{ \pa X^a}~,  \\
& [\cD_{\bar A}  ,  \cQ_{\bar B} \} ~=~0 ~.
\end{aligned}
\ee
We can also derive these anti-commutation relations
\bea
\{ \cD^{\a}, \bar{\cD}^{\ad} \} = 2 \ri (\tilde{\s}^a)^{\ad \a} \frac{\pa}{\pa X^a}~, \qquad \{ {\cQ}^{\a}, \bar{\cQ}^{\ad} \} = -2 \ri\, (\tilde{\s}^a)^{\ad \a} \frac{\pa}{\pa X^a}~.
\eea

Let $\F^{\cA_1}, \J^{\cA_2}$ and $\Pi^{\cA_3}$ be primary superfields with weights $(q_1, \bar{q}_1),(q_2, \bar{q}_2)$ and $(q_3, \bar{q}_3)$ respectively. Then, the three-point correlation function may be constructed using the general expression \cite{OsbornN1, Park1, Park}:
\bea \label{3ptgen}
&&\langle
\F^{\cA_1} (z_1) \, \J^{\cA_2}(z_2)\,  \Pi^{\cA_3}(z_3)
\rangle 
=\frac{ 
\cI^{\cA_1 \cB_1} ({{x}_{1 \bar{3}}}, {{x}_{3 \bar{1}}})\,
\cI^{\cA_2 \cB_2} ({{x}_{2 \bar{3}}}, {{x}_{3 \bar{2}}})\,
}
{ 
(x_{\bar{1}3}{}^2)^{\bar{q}_1} (x_{\bar{3}1}{}^2)^{q_1} 
(x_{\bar{2}3}{}^2)^{\bar{q}_2} (x_{\bar{3}2}{}^2)^{q_2}
}
H_{\cB_1 \cB_2}{}^{\cA_3} (\fn3)~,~~~~~~~~
\eea 
where the functional form of the tensor $H_{\cB_1 \cB_2}{}^{\cA_3}$ is highly constrained by the superconformal symmetry as follows:
\begin{itemize}

\item[(i)] It possesses the homogeneity property
\be
\begin{aligned}
& H_{\cB_1 \cB_2}{}^{\cA_3} ( \l \bar{\l}\, {X},
\l\, {\Q}, \bar{\l} \bar {\Q}) =
\l^{2a} \bar{\l}^{2\bar{a}}
H_{\cB_1 \cB_2}{}^{\cA_3}( {X}, {\Q}, \bar {\Q})~, \\
& a- 2\bar{a} = \bar{q}_1 + \bar{q}_2 -q_3~,\qquad
\bar{a} - 2a = q_1 + q_2 - \bar{q}_3~.
\end{aligned}
\ee
This condition guarantees that the correlation function has the correct transformation law under the superconformal group. 
By construction, eq.~\eqref{3ptgen} has the correct transformation properties at the points $z_1$ and $z_2$. 
The above homogeneity property implies that it also transforms correctly at $z_3$. The tensor $ H_{\cB_1 \cB_2}{}^{\cA_3}$ has dimension $(a + \bar{a})$.

\item[(ii)]If any of the superfields $\F$, $\J$ and $\P$ obey differential equations (e.g. conservation laws for conserved current multiplets), then $ H_{\cB_1 \cB_2}{}^{\cA_3}$
is constrained by certain differential equations too. The latter may be derived using \eqref{Cderivs}.

\item[(iii)] If any (or all) of
the superfields $\F$, $\J$ and $\P$ coincide,
then $ H_{\cB_1 \cB_2}{}^{\cA_3}$
obeys additional constraints, the so-called ``point-switch symmetries". These are consequences of the symmetry
under permutations of superspace points. As an example,
\be
\langle \Phi^{{\cal A}}(z_1) \Phi^{{\cal B}}(z_2)
\P^{{\cal C}}(z_3) \rangle =
(-1)^{\epsilon(\Phi)}
\langle \Phi^{{\cal B}}(z_2) \Phi^{{\cal A}}(z_1)
\P^{{\cal C}}(z_3) \rangle~,
\ee
where $\epsilon(\Phi)$ denotes the Grassmann parity of $\Phi^{{\cal A}}$. Note that under permutations of any two superspace points, the three-point building blocks transform as
\bsubeq
\bea
			{X}_{3 \, \a \ad} &\stackrel{1 \leftrightarrow 2}{\longrightarrow} - \bar{X}_{3 \, \a \ad} \, , \hspace{10mm} {\Q}_{3 \, \a} \stackrel{1 \leftrightarrow 2}{\longrightarrow} - {\Q}_{3 \, \a} \, , \label{pt12} \\[2mm]
			{X}_{3 \, \a \ad} &\stackrel{2 \leftrightarrow 3}{\longrightarrow} - \bar{X}_{2 \, \a \ad} \, , \hspace{10mm} {\Q}_{3 \, \a} \stackrel{2 \leftrightarrow 3}{\longrightarrow} - {\Q}_{2 \, \a} \, , \label{pt23} \\[2mm]
			{X}_{3 \, \a \ad} &\stackrel{1 \leftrightarrow 3}{\longrightarrow} - \bar{X}_{1 \, \a \ad} \, , \hspace{10mm} {\Q}_{3 \, \a} \stackrel{1 \leftrightarrow 3}{\longrightarrow} - {\Q}_{1 \, \a} \, . \label{pt13}
\eea
\esubeq
\end{itemize}
The above conditions fix the functional form of $ H_{\cB_1 \cB_2}{}^{\cA_3}$ (and, therefore, the three-point function under consideration) up to a few arbitrary constants.

A few comments are in order regarding the three-point functions of conserved current multiplets. It is worth pointing out that, depending on the exact way in which one constructs the general expression \eqref{3ptgen}, it can be impractical to impose conservation equations on one of the three superfields due to a lack of useful identities such as \eqref{Cderivs}. To illustrate this, let us go back to eq. \eqref{3ptgen}:
\bea 
&&\langle
\F^{\cA_1} (z_1) \, \J^{\cA_2}(z_2)\,  \Pi^{\cA_3}(z_3)
\rangle 
=\frac{ 
\cI^{\cA_1 \cB_1} ({{x}_{1 \bar{3}}}, {{x}_{3 \bar{1}}})\,
\cI^{\cA_2 \cB_2} ({{x}_{2 \bar{3}}}, {{x}_{3 \bar{2}}})\,
}
{ 
(x_{\bar{1}3}{}^2)^{\bar{q}_1} (x_{\bar{3}1}{}^2)^{q_1} 
(x_{\bar{2}3}{}^2)^{\bar{q}_2} (x_{\bar{3}2}{}^2)^{q_2}
}
H_{\cB_1 \cB_2}{}^{\cA_3} (\fn3)~,~~~~~~~~
\label{3ptgen-new}
\eea 
All information about this correlation function is encoded in the tensor $H$; however, this particular formulation prevents us from imposing conservation on $\P$ in a straightforward way. A way out is to rearrange the correlator with $\P$, say, at the second point:
\bea 
&&\langle
\J^{\cA_2}(z_2)\,  \Pi^{\cA_3}(z_3) \, \F^{\cA_1} (z_1)\,  
\rangle 
=\frac{ 
\cI^{\cA_2 \cB_2} ({{x}_{2 \bar{1}}}, {{x}_{1 \bar{2}}})\,
\cI^{\cA_3 \cB_3} ({{x}_{1 \bar{3}}}, {{x}_{3 \bar{1}}})\,
}
{ 
(x_{\bar{2}1}{}^2)^{\bar{q}_2} (x_{\bar{1}2}{}^2)^{q_2} 
(x_{\bar{3}1}{}^2)^{\bar{q}_3} (x_{\bar{1}3}{}^2)^{q_3}
}
\tilde{H}_{\cB_2 \cB_3}{}^{\cA_1} (X_1, \Q_1, \bar{\Q}_1)~.~~~~~~~~
\label{3ptgen-231}
\eea 
Here, all information about the correlator is now encoded in the tensor $\tilde{H}$, which is a completely different solution compared to $H$. 
Thus, we require a simple equation relating the tensors $H$ and $\tilde{H}$, which corresponds to different representations of the same correlation function.
Indeed, once $\tilde{H}$ is obtained, we can then easily impose conservation on $\Pi$ as if it were located at the ``second point'', with the aid of identities analogous 
to \eqref{Cderiv-2a} and \eqref{Cderiv-2b}. However, as we will see in the next sections, this transformation proves to be complicated for correlators of higher-spin primary 
operators due to the proliferation of tensor/spinor indices. In section \ref{section3}, we will develop an index-free approach to study the correlator involving three insertions of the 
higher-spin supercurrent $J_{\a(r) \ad(r)}$ and derive the explicit formula relating $\tilde{H}$ and $H$.

 
\section{General formalism}
\label{section3}


In this section we will develop the general formalism to derive all the necessary constraints on the three-point function containing three insertions of 
the higher-spin supercurrent $J_{(m_1 \dots m_r)} \sim  J_{\a (r) \ad (r)}$.
The ansatz for the correlator consistent with the expression \eqref{3ptgen} is given by
\bea \label{3HS-currents}
&&\langle
J_{\a(r_1) \ad(r_1)} (z_1) \, J'{}_{\b(r_2) \bd(r_2)}(z_2)\,  J''{}_{\g(r_3) \gd(r_3)}(z_3)
\rangle \non\\
&&=\frac{1}{k_1} \cI_{\a(r_1) \dot{\mu}(r_1)} (x_{1 \bar{3}}) \cI_{{\mu}(r_1) \ad(r_1)} (x_{3 \bar{1}}) \cI_{\b(r_2) \dot{\l}(r_2)} (x_{2 \bar{3}}) \cI_{{\l}(r_2) \bd(r_2)} (x_{3 \bar{2}})\non\\
&& \qquad \times H^{\dot{\mu}(r_1) \mu(r_1), \, \dot{\l}(r_2) \l(r_2)}{}_{\g(r_3) \dot{\g}(r_3)} (\fn3)~,~~~~~~~~
\eea 
with $k_1 := (x_{\bar{1} 3} x_{\bar{3} 1})^{r_1+2} (x_{\bar{2} 3} x_{\bar{3} 2})^{r_2+2}$. We recall that $J_{\a(r) \ad(r)}$ is primary real with weights $(q, \bar{q}) = (\frac{r+2}{2}, \frac{r+2}{2})$ and scaling dimension $r+2$. We first assume that all the operators are of different spins; hence, we will not impose any point-switch symmetries. 
The correlator \eqref{3HS-currents} is thus subject to the following constraints:
\begin{enumerate}
	\item[\textbf{(i)}] \textbf{Homogeneity:} The tensor $H_{\mu(r_1) \dot{\mu}(r_1), \, \l(r_2) \dot{\l}(r_2), \, \g(r_3) \gd(r_3)}$ has the scaling property
	\bea
	&&H_{\mu(r_1) \dot{\mu}(r_1), \, \l(r_2) \dot{\l}(r_2), \, \g(r_3) \gd(r_3)} (\L \bar{\L} X, \L \Q, \bar{\L} \bar{\Q}) \non\\
	&=& (\L \bar{\L})^{r_3-(r_1+r_2+2)} H_{\mu(r_1) \dot{\mu}(r_1), \, \l(r_2) \dot{\l}(r_2), \, \g(r_3) \gd(r_3)} (\fxq)~, 	
	\eea
and, hence, its dimension is $r_3-(r_1+r_2+2)$. This ensures that the correlator transforms correctly under scale transformations.
	\item[\textbf{(ii)}] \textbf{Conservation:} The conservation of $J_{\a(r) \ad(r)}$ at $z_{1}$ and $z_{2}$ imply
\bsubeq
\bea
D^{\a_1}_{(1)} \corr1 &=& 0~, \\
\bar D^{\ad_1}_{(1)}  \corr1 &=& 0~, \\
D^{\b_1}_{(2)}  \corr1 &=& 0~, \\
\bar D^{\bd_1}_{(2)} \corr1 &=& 0~.
\eea
\esubeq
With the use of identities \eqref{Cderivs}, these requirements are translated to the following differential constraints on $H$:
\bsubeq \label{ceq}
\bea
&&\bar{\cD}^{\dot{\d}} H_{\mu(r_1) \dot{\d}\dot{\mu}(r_1-1), \, \l(r_2) \dot{\l}(r_2), \, \g(r_3) \gd(r_3)} = 0~,
\label{ceq1}\\
&&{\cD}^{{\d}} H_{\d \mu(r_1-1) \dot{\mu}(r_1), \, \l(r_2) \dot{\l}(r_2), \, \g(r_3) \gd(r_3)} = 0~, \label{ceq2}\\
&&\bar{\cQ}^{\dot{\d}} H_{\mu(r_1) \dot{\mu}(r_1), \, \l(r_2) \dot{\d} \dot{\l}(r_2-1), \, \g(r_3) \gd(r_3)} = 0~, \label{ceq3}\\
&&\cQ^{\d}H_{\mu(r_1) \dot{\mu}(r_1), \, \d \l(r_2-1) \dot{\l}(r_2), \, \g(r_3) \gd(r_3)} = 0~.\label{ceq4}
\eea
\esubeq
There are further constraints arising from the conservation at $z_3$:
	\bea
	D^{\g_1}_{(3)}  \corr1 &=& 0~, \\
	\bar D^{\gd_1}_{(3)} \corr1 &=& 0~.	
	\eea
These constraints are non-trivial to impose as there are no identities analogous to \eqref{Cderivs} which allow the spinor derivatives acting on $z_{3}$ to pass through the prefactor of \eqref{3HS-currents}. We thus employ the procedure outlined at the end of subsection 2.2, where in eq.\,\eqref{3ptgen-231} we reformulated the ansatz in terms of $\tilde{H}$ and with $\P$ at the second point. The details of this will be illustrated in the next subsection.

\item[\textbf{(iii)}] \textbf{Reality:} Since the higher-spin supercurrent $J_{\a(r) \ad(r)}$ is a real superfield, the reality condition on the correlator leads to the following constraint on the tensor $H$:
\bea
H_{\mu(r_1) \dot{\mu}(r_1), \, \l(r_2) \dot{\l}(r_2), \, \g(r_3) \gd(r_3)} (\fxq) = \bar{H}_{\mu(r_1) \dot{\mu}(r_1), \, \l(r_2) \dot{\l}(r_2), \, \g(r_3) \gd(r_3)} (X, \Q, \bar{\Q})~,~~~~
\eea
\end{enumerate}
where $\bar{H}(\fxq)$ is the conjugate of $H(\fxq)$ (with indices suppressed). More precisely, we introduce the following definitions.  Suppose $H(\fxq)$ is composed out of a finite basis of linearly independent tensor structures $T_{i}(\fxq)$, that is $H(\fxq) = \sum_{i} c_{i} T_{i}(\fxq)$, with $c_{i}$ constant complex parameters. Define the conjugate of $H$ by
\bea
\bar{H}(\fxq) &=& \sum_{i} \bar{c}_{i} \bar{T}_{i} (\fxq) 
\non\\
&=& \sum_{i} \bar{c}_{i} T_{i}(\bar{X}, \bar \Q, \Q) ~.
\label{Hbar-cc} 
\eea
The problem of computing higher-spin correlator \eqref{3HS-currents} is thus reduced to determining the most general form of $H(\fxq)$ subject to the above constraints. The first observation is that since $H$ is Grassmann even, it must be an even function of $\Q$ and $\bar{\Q}$. 
We can then write a general expansion for $H$ as follows (to recall, $P_{\dot{\d} \d} =- 4 \Theta_{\d} \bar \Theta_{\dot{\d}}$):
\bea
H(\fxq) &=& F(X) -\hf P^{\dot{\d} \d} G_{\d \dot{\d}} (X) \non\\
&& + A^{(1)} (X) \,\Q^2 + A^{(2)} (X) \,\bar{\Q}^2 + A^{(3)} (X)\, \Q^2 \bar{\Q}^2~.
\eea
Now, constraints \eqref{ceq1} and \eqref{ceq4} from conservation imply the following conditions
\bea
\bar{\cD}^2 H(\fxq)=0~, \qquad
{\cQ}^2 H(\fxq)=0~,
\eea
which lead to
\bea
A^{(1)} (X) = A^{(2)} (X)= A^{(3)} (X) = 0~.
\eea
As a result, the general solution for $H(\fxq)$ can always be presented in the form
\be
\begin{aligned}
&H_{\a(r_1) \ad(r_1), \, \b(r_2) \bd(r_2), \, \g(r_3) \gd(r_3)} (\fxq) \\
&=F_{\a(r_1) \ad(r_1), \, \b(r_2) \bd(r_2), \, \g(r_3) \gd(r_3)} ({X})\\
&\quad -\hf P^{\dot{\d} \d} G_{\d \dot{\d},\,\a(r_1) \ad(r_1), \, \b(r_2) \bd(r_2), \, \g(r_3) \gd(r_3)} ({X})~.~~~~~~~
\end{aligned}
\label{H-FG}
\ee
Our task is to find the appropriate tensorial structures for $F$ and $G$, consistent with homogeneity, conservation and reality constraints. 
This is a technically challenging problem due to the proliferation of spinor indices. It is advantageous to develop an index-free approach which allows us to systematically write the ansatz for $H$, which is an extension of the non-supersymmetric generating function formalism recently proposed in \cite{BS22}. In this work, we provide a formalism to efficiently impose all the required (differential) constraints on the three-point function of the conserved  higher-spin supercurrent \eqref{3HS-currents}.
These constraints can then be efficiently implemented and solved computationally by \textit{Mathematica}.

\subsection{Generating function}
We begin by introducing a method that allows us to encode symmetric traceless tensors by polynomials obtained by contracting the tensor with sets of commuting auxiliary spinors.  Let $(u^{\a}, \bar{u}^{\ad})$ be a set of commuting auxiliary spinors satisfying, by construction, $u^2 = u^{\a}u_{\a} = 0;\, \bar{u}^{2} =\bar{u}^{\ad} \bar{u}_{\ad} = 0$. This is, in fact, equivalent to defining a null vector $U^{a}$,
\bea
U^a = -\hf (\tilde{\s}^a)^{\ad \a} U_{\a \ad}~, \qquad U_{\a \ad} = u_{\a} \bar{u}_{\ad}~,
\eea
such that $U^2 = U^a U_a = 0$. 
Extending this construction, we define
\bsubeq
\bea
\mathbf{U}^{\ad(r_1) \a(r_1)} &=& u^{\a_1} \dots u^{\a_{r_1}} \bar{u}^{\ad_1} \dots \bar{u}^{\ad_{r_1}}~, \\
\mathbf{V}^{\bd(r_2) \b(r_2) } &=& v^{\b_1} \dots v^{\b_{r_2}} \bar{v}^{\bd_1} \dots \bar{v}^{\bd_{r_2}}~,\\
\mathbf{W}^{\gd(r_3) \g(r_3) } &=& w^{\g_1} \dots w^{\g_{r_3}} \bar{w}^{\gd_1} \dots \bar{w}^{\gd_{r_3}}~.
\eea
\esubeq
Here the spinors satisfy $u^2 = \bar{u}^2 = 0,~ v^2 = \bar{v}^2 =0,~  w^2= \bar{w}^2=0$. 
We then represent $H_{\a(r_1) \ad(r_1), \, \b(r_2) \bd(r_2), \, \g(r_3) \gd(r_3)} (\fxq)$ in terms of the following generating polynomial
\bea
&&H(X,\Q, \bar{\Q}; u, \bar{u}, v, \bar{v}, w, \bar{w}) \non\\
&=& \mathbf{U}^{\ad(r_1) \a(r_1) } \mathbf{V}^{\bd(r_2) \b(r_2) } \mathbf{W}^{\gd(r_3) \g(r_3) } H_{\a(r_1) \ad(r_1), \, \b(r_2) \bd(r_2), \, \g(r_3) \gd(r_3)} (\fxq)~. \label{polynomial-H}
\eea
The tensor $H$ can be extracted from the generating polynomial by acting on it with partial derivatives
\bea
&&H_{\a(r_1) \ad(r_1), \, \b(r_2) \bd(r_2), \, \g(r_3) \gd(r_3)} (\fxq)  \non\\
&&= 
\frac{\pa}{ \pa \mathbf{U}^{\ad(r_1) \a(r_1)} } \frac{\pa}{ \pa \mathbf{V}^{\bd(r_2) \b(r_2)}} \frac{\pa}{ \pa  \mathbf{W}^{\gd(r_3) \g(r_3)}} \, H(\fxq; u, \bar{u}, v, \bar{v}, w, \bar{w} )\, .
\eea
In the above, the partial derivative operators are defined by
\bsubeq
\bea
		\frac{\pa}{\pa \mathbf{U}^{\a(r_{1}) \ad(r_{1})} } = \frac{1}{r_{1}!r_{1}!} \frac{\pa}{\pa u^{\a_{1}} } \dots \frac{\pa}{\pa u^{\a_{r_{1}}}} \frac{\pa}{\pa \bar{u}^{\ad_{1}}} \dots \frac{\pa }{\pa \bar{u}^{\ad_{r_{1}}}} \, , \\
		\frac{\pa}{\pa \mathbf{V}^{\b(r_{2}) \bd(r_{2})} } = \frac{1}{r_{2}!r_{2}!} \frac{\pa}{\pa v^{\b_{1}} } \dots \frac{\pa}{\pa v^{\b_{r_{2}}}} \frac{\pa}{\pa \bar{v}^{\bd_{1}}} \dots \frac{\pa }{\pa \bar{v}^{\bd_{r_{2}}}} \, , \\
		\frac{\pa}{\pa \mathbf{W}^{\g(r_{3}) \gd(r_{3})} } = \frac{1}{r_{3}!r_{3}!} \frac{\pa}{\pa w^{\g_{1}} } \dots \frac{\pa}{\pa w^{\g_{r_{3}}}} \frac{\pa}{\pa \bar{w}^{\gd_{1}}} \dots \frac{\pa }{\pa \bar{w}^{\gd_{r_{3}}}} \, . 
\eea
\esubeq
From the general form \eqref{H-FG}, we find that
\bea
&&H(X,\Q, \bar{\Q}; u, \bar{u}, v, \bar{v}, w, \bar{w}) \non\\
&=& F({X}; u, \bar{u}, v, \bar{v}, w, \bar{w}) -\hf P^{\ad \a}G_{\a \ad}({X}; u, \bar{u}, v, \bar{v}, w, \bar{w})~.
\eea
Hence, it suffices to work with polynomials $F({X}; u, \bar{u}, v, \bar{v}, w, \bar{w})$ and $G_{\a \ad}({X}; u, \bar{u}, v, \bar{v}, w, \bar{w})$, which are functions of a \textit{single, bosonic} covariant building block $X$, and the auxiliary commuting spinors $u, \bar{u}, v, \bar{v}, w, \bar{w}$. 
To ensure that \eqref{polynomial-H} is satisfied, the polynomials  $F({X}; u, \bar{u}, v, \bar{v}, w, \bar{w})$ and $G_{\a \ad}({X}; u, \bar{u}, v, \bar{v}, w, \bar{w})$ must be homogeneous of degree $r_1$ in $u, \bar{u}$, degree $r_2$ in $v, \bar{v}$ and degree $r_3$ in $w, \bar{w}$ (equivalently, we can view them as homogeneous polynomials of degree $r_1$, $r_2$ and $r_3$ in $U$, $V$ and $W$, respectively).

Next, we define the invariants which serve as the building blocks for constructing solutions for $F({X}; u, \bar{u}, v, \bar{v}, w, \bar{w})$ and $G_{\a \ad}({X}; u, \bar{u}, v, \bar{v}, w, \bar{w})$. Since $H$ has a scale dimension $d = r_3-(r_1+r_2+2)$, we can then write
\bea
F({X}; u, \bar{u}, v, \bar{v}, w, \bar{w}) = \frac{1}{X^{-d}} \cF(\hat{X};u, \bar{u}, v, \bar{v}, w, \bar{w} ).
\eea
Here the polynomial $\cF(\hat{X};u, \bar{u}, v, \bar{v}, w, \bar{w} )$ is now homogeneous of degree 0 in $X$ and is constructed out of scalar combinations of $\hat{X}$, along with the auxiliary spinors $U$, $V$ and $W$ with the appropriate homogeneity. We choose the following basis for the invariants:
\bsubeq \label{basis-F}
\bea
&&UV = -\hf (u\cdot v) (\bar{u} \cdot \bar{v})~, \, \, \,UW = -\hf (u \cdot w) (\bar{u} \cdot \bar{w})~, \, \,\,VW = -\hf (v \cdot w) (\bar{v} \cdot \bar{w})~,~~~~~~~~~~~ \\
&&U \hat{X} = U^a \hat{X}_{a}= -\hf \hat{X}_{(u, \bar{u})}~, \qquad V \hat{X} = -\hf \hat{X}_{(v, \bar{v})}~, \quad \quad \,\,\,W \hat{X} = -\hf \hat{X}_{(w, \bar{w})}~,~~~~~~~~~  \\
&& \cJ = \e_{abcd}\hat{X}^a U^b V^c W^d = \frac{\ri}{4} \big[ \hat{X}_{(u, \bar{w})} (v \cdot w) (\bar{u} \cdot \bar{v}) - \hat{X}_{(w, \bar{u})} (u \cdot v) (\bar{v} \cdot \bar{w})
\big]~,
\eea
\esubeq
where we have defined $(u \cdot v) (\bar{u} \cdot \bar{v}) = (u^{\a} v_{\a}) (\bar{u}^{\ad} \bar{v}_{\ad})$ and $\hat{X}_{(u, \bar{u})} = u^{\a} \hat{X}_{\a \ad} \bar{u}^{\ad}$. 
The polynomial $\cF$ can be expressed as
\bsubeq\label{expansion-F-scalar}
\bea
&&\cF(\hat{X};u, \bar{u}, v, \bar{v}, w, \bar{w} ) = \cF_1(\hat{X};u, \bar{u}, v, \bar{v}, w, \bar{w} )+ \cF_2(\hat{X};u, \bar{u}, v, \bar{v}, w, \bar{w} )~,
\eea
where we define
\bea
&&\cF_1(\hat{X};u, \bar{u}, v, \bar{v}, w, \bar{w} ) = \sum_{p_i} A(p_i) (UV)^{p_1} (UW)^{p_2} (VW)^{p_3} \non\\
&&\hspace{5cm} \times~(U \hat{X})^{p_4}(V \hat{X})^{p_5} (W\hat{X})^{p_6}~,\\
&&\cF_2(\hat{X};u, \bar{u}, v, \bar{v}, w, \bar{w} ) = \sum_{q_i} A'(q_i) (UV)^{q_1} (UW)^{q_2} (VW)^{q_3} \non\\
&&\hspace{5cm} \times~(U \hat{X})^{q_4}(V \hat{X})^{q_5} (W\hat{X})^{q_6} \cJ~,
\eea
\esubeq
for a set of complex constants $A(p_i), A'(q_i)$. 
In the above we can only have at most a linear term in $\cJ$, thanks to the relation $\e_{abcd} \e^{a'b'c'd'} = \d^{a'}_{[a} \dots \d^{d'}_{d]}$. 
The reason for separating the $\cJ$-independent and $\cJ$-dependent terms  will be clear later.
Due to the homogeneity constraints on $\cF$, here $p_i$ and $q_i$ are non-negative integers which solve the following linear system of equations:
\bsubeq
\bea
p_1 + p_2 + p_4 &=& q_1 + q_2 + q_4 +1 = r_1~,\\
p_1 + p_3 + p_5 &=& q_1 + q_3 + q_5 + 1 = r_2~,\\
p_2 + p_3 + p_6 &=& q_2 + q_3 + q_6 + 1 = r_3~.
\eea
\esubeq

In constructing a basis of invariants for  $G_{\a \ad}({X}; u, \bar{u}, v, \bar{v}, w, \bar{w})$, let us write
\bea
G_{\a \ad}({X}; u, \bar{u}, v, \bar{v}, w, \bar{w}) = \frac{1}{X^{1-d}} \cG_{\a \ad}(\hat{X};u, \bar{u}, v, \bar{v}, w, \bar{w} ).
\eea
Now, we observe that $P^a \cG_{a} (\hat{X};u, \bar{u}, v, \bar{v}, w, \bar{w} )$ is constructed out of scalar combinations of $\hat{X}$, along with the auxiliary spinors $U$, $V$ and $W$ with the appropriate homogeneity. We can then construct
\bsubeq
\bea
&&P^a \cG_{a} (\hat{X};u, \bar{u}, v, \bar{v}, w, \bar{w} ) 
= 
P^a \bigg[
U_{a} \,\bm{\cG}_{1} (\hat{X};u, \bar{u}, v, \bar{v}, w, \bar{w} )+\hat{X}_{a} \,\bm{\cG}_{2} (\hat{X};u, \bar{u}, v, \bar{v}, w, \bar{w}) \non\\
&&\hspace{3cm}+\,{V}_{a}\, \bm{\cG}_{3} (\hat{X};u, \bar{u}, v, \bar{v}, w, \bar{w})+ {W}_{a} \,\bm{\cG}_{4} (\hat{X};u, \bar{u}, v, \bar{v}, w, \bar{w}) \non\\
&&\hspace{3cm}+\,{\cZ}_{1a}\, \bm{\cG}_{5} (\hat{X};u, \bar{u}, v, \bar{v}, w, \bar{w}) + {\cZ}_{2a}\bm{\cG}_{6} (\hat{X};u, \bar{u}, v, \bar{v}, w, \bar{w})\non\\
&&\hspace{3cm}+\,{\cZ}_{3a} \,\bm{\cG}_{7} (\hat{X};u, \bar{u}, v, \bar{v}, w, \bar{w})+ {\cZ}_{4a} \bm{\cG}_{8} (\hat{X};u, \bar{u}, v, \bar{v}, w, \bar{w}) \bigg]~. \label{expansion-G-vector}
\eea
where we have defined
\bea
&&\cZ_{1a} = \e_{abcd}\hat{X}^{b} U^c V^d~, \qquad \quad \cZ_{2a} = \e_{abcd}\hat{X}^{b} U^c W^d~,\non\\
&&\cZ_{3a} = \e_{abcd}\hat{X}^{b} V^c W^d~, \qquad \quad \cZ_{4a} = \e_{abcd}{U}^{b} V^c W^d~.
\eea
\esubeq
We may express the expansion \eqref{expansion-G-vector} in spinor notation using $P^a \cG_{a} (\hat{X};u, \bar{u}, v, \bar{v}, w, \bar{w} ) = -\hf P^{\ad \a} (\s^a)_{\a \ad} \cG_{a} (\hat{X};u, \bar{u}, v, \bar{v}, w, \bar{w} ) $. This gives
\bsubeq \label{basis-G}
\bea
\cG_{\a \ad} (\hat{X};u, \bar{u}, v, \bar{v}, w, \bar{w} ) &=& (\s^a)_{\a \ad} \cG_{a} (\hat{X};u, \bar{u}, v, \bar{v}, w, \bar{w} )  \non\\
&=& \sum_{I=1}^8 \cK_{I,\,\a \ad} \bm{\cG}_{I} (\hat{X};u, \bar{u}, v, \bar{v}, w, \bar{w} ) ~,
\eea
where the $\cK_{I,\,\a \ad}$ structures take the form
\bea
\cK_{1,\,\a \ad} &=& U_{\a \ad} = u_{\a} \bar{u}_{\ad}~, \non\\
\cK_{2,\,\a \ad} &=& \hat{X}_{\a \ad}~, \non\\
\cK_{3,\,\a \ad} &=& V_{\a \ad} = v_{\a} \bar{v}_{\ad}~, \non\\
\cK_{4,\,\a \ad} &=& W_{\a \ad} = w_{\a} \bar{w}_{\ad}~, \non\\
\cK_{5,\,\a \ad} &=& \cZ_{1, \, \a \ad} 
= \frac{\ri}{2} \Big[ \hat{X}_{\a \bd} \bar{v}_{\ad} \bar{u}^{\bd}(u \cdot v) - \hat{X}_{\b \ad} {v}_{\a}{u}^{\b} (\bar{u} \cdot \bar{v})\Big]~, \\
\cK_{6,\,\a \ad} &=& \cZ_{2, \, \a \ad} 
= \frac{\ri}{2} \Big[ \hat{X}_{\a \bd} \bar{w}_{\ad} \bar{u}^{\bd}(u \cdot w) - \hat{X}_{\b \ad} {w}_{\a}{u}^{\b} (\bar{u} \cdot \bar{w})\Big]~, \non\\
\cK_{7,\,\a \ad} &=& \cZ_{3, \, \a \ad} 
= \frac{\ri}{2} \Big[ \hat{X}_{\a \bd} \bar{w}_{\ad} \bar{v}^{\bd}(v \cdot w) - \hat{X}_{\b \ad} {w}_{\a}{v}^{\b} (\bar{v} \cdot \bar{w})\Big]~, \non\\
\cK_{8,\,\a \ad} &=& \cZ_{4, \, \a \ad} 
= \frac{\ri}{2} \Big[ w_{\a} \bar{u}_{\ad}(u \cdot v) (\bar{v} \cdot \bar{w}) - u_{\a} \bar{w}_{\ad}(v \cdot w) (\bar{u} \cdot \bar{v})\Big]~.\non
\eea
\esubeq
It should be noted, however, that not all of the $\cK$-structures are linearly independent. For instance, we have that
\bea
v_{\a} \bar{v}_{\ad} \cJ + (V W) \cZ_{1\, \a \ad} + (U V) \cZ_{3\, \a \ad} - \cZ_{4\, \a \ad}(V \hat{X}) = 0~.
\eea 
As will be shown in section 4, there will be more linear dependence relations. 
Now that $\bm{\cG}_{I} (\hat{X};u, \bar{u}, v, \bar{v}, w, \bar{w} )$ are scalars, they can be constructed in an analogous way as in $\cF_1$ and $\cF_2$:
\bea
&&\bm{\cG}_{I} (\hat{X};u, \bar{u}, v, \bar{v}, w, \bar{w} ) = 
\sum_{p_i} B_I(p_i) (UV)^{p_1} (UW)^{p_2} (VW)^{p_3}
(U\hat{X})^{p_4}(V \hat{X})^{p_5} (W\hat{X})^{p_6}\non\\
&&\hspace{3cm} +\sum_{q_i} B'_I(q_i) (UV)^{q_1} (UW)^{q_2} (VW)^{q_3}(U \hat{X})^{q_4}(V \hat{X})^{q_5} (W\hat{X})^{q_6} \cJ~.~~~~~~~~ \label{exp-G-scalar}
\eea

Let us now recast the conservation and reality constraints \textbf{(ii)} and \textbf{(iii)} in terms of $F$ and $G$. Since $\bar{\cD}_{\ad} = -\pa /\pa \bar{\Q}^{\ad}$, the first differential constraint \eqref{ceq1} simply turns into
\bea
\ve^{\ad \bd} \frac{\pa}{\pa \bar{u}^{\bd}} G_{\a \ad}({X}; u, \bar{u}, v, \bar{v}, w, \bar{w}) = 0~.
\eea
In the second equation \eqref{ceq2}, note that $\cD_{\a} = {\pa}/{ \pa {\Q}^{\a} }
-2{\rm i}\, \bar{\Q}^\ad \pa_{\a \ad}$\,, with $\pa_{\a \ad} = (\s^a)_{\a \ad}{\pa }/{\pa X^a}$.  Thus, this gives rise to two conditions:
\bsubeq
\bea
&&\pa^{\ad \b} \frac{\pa}{\pa u^{\b}} G_{\a \ad} (X; u, \bar{u}, v, \bar{v}, w, \bar{w}) = 0~, \\
&&\ve^{\a \b} \frac{\pa}{\pa u^{\b}} \Big[ G_{\a \ad}(X; u, \bar{u}, v, \bar{v}, w, \bar{w})- \ri  \pa_{\a \ad} F(X; u, \bar{u}, v, \bar{v}, w, \bar{w}) \Big] = 0~.
\eea
\esubeq
In a similar fashion, it is not hard to see that \eqref{ceq3} is equivalent to 
\bsubeq \label{ceq2-short}
\bea
&&\pa^{\bd \a} \frac{\pa}{\pa \bar{v}^{\bd}} G_{\a \ad} (X; u, \bar{u}, v, \bar{v}, w, \bar{w}) = 0~, \label{ceq2-short-a} \\
&&\ve^{\ad \bd} \frac{\pa}{\pa \bar{v}^{\bd}} \Big[ G_{\a \ad}(X; u, \bar{u}, v, \bar{v}, w, \bar{w})- \ri  \pa_{\a \ad} F(X; u, \bar{u}, v, \bar{v}, w, \bar{w}) \Big] = 0~,\label{ceq2-short-b}
\eea
\esubeq
while the last equation \eqref{ceq4} implies that
\bea
\ve^{\a \b} \frac{\pa}{\pa {v}^{\b}} G_{\a \ad}({X}; u, \bar{u}, v, \bar{v}, w, \bar{w}) = 0~.
\eea
We also have a nice consistency check which follows directly from eqs.\,\eqref{ceq2-short-a} and \eqref{ceq2-short-b}:
\bea
\Box F({X}; u, \bar{u}, v, \bar{v}, w, \bar{w}) - \ri \pa^{\ad \ad} G_{\a \ad} ({X}; u, \bar{u}, v, \bar{v}, w, \bar{w}) = 0~, \quad \Box = -\hf \pa^{\ad \a} \pa_{\a \ad}~.~~~~~
\label{consistency}
\eea
As for the reality constraint, note that \eqref{Hbar-cc} allows us to express the conjugate of $H$ as
\bea
&&\bar{H}(X,\Q, \bar{\Q}; u, \bar{u}, v, \bar{v}, w, \bar{w})\non\\
&=& F(\bar{X}; u, \bar{u}, v, \bar{v}, w, \bar{w}; \,\bar{a}_i)- \hf P^{\ad \a} G_{\a \ad}(\bar{X}; u, \bar{u}, v, \bar{v}, w, \bar{w};\, \bar{b}_i)~.
\label{Hbar-FG}
\eea
Here $a_i$ and $b_i$ are constant complex parameters of the linearly independent tensor structures in $F(X; u, \bar{u}, v, \bar{v}, w, \bar{w})$ and $G_{\a \ad}(X; u, \bar{u}, v, \bar{v}, w, \bar{w})$, respectively.\footnote{These parameters correspond to $A(p_i), A'(q_i), B_I(p_i), B'_I(q_i)$ in eqs.~\eqref{expansion-F-scalar} and \eqref{exp-G-scalar}.} With the definitions \eqref{Xbar-def} and the fact that $\Q P^2 = \bar{\Q} P^2 = 0$, we can apply Taylor's expansion on the right-hand side of \eqref{Hbar-FG} and arrive at
\be
\begin{aligned}
&\bar{H}(X,\Q, \bar{\Q}; u, \bar{u}, v, \bar{v}, w, \bar{w})\\
&= F(X;  u, \bar{u}, v, \bar{v}, w, \bar{w};\,\bar{a}_i) \\
&- \frac{1}{2} P^{\ad \a} \Big[ \ri \pa_{\a \ad} F(X; u, \bar{u}, v, \bar{v}, w, \bar{w}; \bar{a}_i) + G_{\a \ad} (X; u, \bar{u}, v, \bar{v}, w, \bar{w}; \,\bar{b}_i) \Big] \\
&- \frac{1}{8}P^2 \Big[ \Box F(X; u, \bar{u}, v, \bar{v}, w, \bar{w}; \,\bar{a}_i)+ \ri \pa^{\ad \a} G_{\a \ad} (X; u, \bar{u}, v, \bar{v}, w, \bar{w}; \,\bar{b}_i) \Big]~.
\label{Hbar-FG-expanded}
\end{aligned}
\ee
The reality constraint reads
\be
\begin{aligned}
&{H}(X,\Q, \bar{\Q}; u, \bar{u}, v, \bar{v}, w, \bar{w}) = \bar{H}(X,\Q, \bar{\Q}; u, \bar{u}, v, \bar{v}, w, \bar{w}) \\
\Longrightarrow ~&F({X}; u, \bar{u}, v, \bar{v}, w, \bar{w}; \,{a}_i)- \hf P^{\ad \ad} G_{\a \ad}({X}; u, \bar{u}, v, \bar{v}, w, \bar{w};\,{b}_i) \\
&= \bar{H}(X,\Q, \bar{\Q}; u, \bar{u}, v, \bar{v}, w, \bar{w})~. \label{re-eq}
\end{aligned}
\ee
Upon substituting \eqref{Hbar-FG-expanded} into the right-hand side of \eqref{re-eq}, it appears that we obtain three conditions:
\bsubeq \label{real-all}
\bea
F({X}; u, \bar{u}, v, \bar{v}, w, \bar{w}; \,{a}_i) &=&
F(X;  u, \bar{u}, v, \bar{v}, w, \bar{w};\,\bar{a}_i)~, \\
G_{\a \ad}({X}; u, \bar{u}, v, \bar{v}, w, \bar{w};\,{b}_i) &=&
G_{\a \ad} (X; u, \bar{u}, v, \bar{v}, w, \bar{w}; \,\bar{b}_i) \non\\ 
&+& \ri \pa_{\a \ad} F(X; u, \bar{u}, v, \bar{v}, w, \bar{w}; \bar{a}_i)~, \label{reality-mixFG}\\
\Box F({X}; u, \bar{u}, v, \bar{v}, w, \bar{w}; \,\bar{a}_i)
&+& \ri \pa^{\ad \a} G_{\a \ad}(X; u, \bar{u}, v, \bar{v}, w, \bar{w}; \,\bar{b}_i) = 0~. \label{reality-boxF}
\eea
\esubeq
However, it is not hard to see that eq.\,\eqref{reality-boxF} follows automatically as a result of imposing \eqref{reality-mixFG} and \eqref{consistency}. Therefore, the reality constraint only demands that the parameters $a_i$ be real and
\bea
G_{\a \ad}({X}; u, \bar{u}, v, \bar{v}, w, \bar{w};\,{b}_i) &=&
G_{\a \ad} (X; u, \bar{u}, v, \bar{v}, w, \bar{w}; \,\bar{b}_i) \non\\ 
&+& \ri \pa_{\a \ad} F(X; u, \bar{u}, v, \bar{v}, w, \bar{w}; {a}_i)~,
\eea
which gives a set of equations relating the imaginary parts of $b_i$ to $a_i$. 

\subsection{Imposing conservation on $z_3$}
We now turn to analysing the conservation condition at $z_3$. In this particular case, we rewrite our correlator as 
\bea
&&\corr1 \non\\
&=& 
\langle
J'{}_{\b(r_2) \bd(r_2)}(z_2)\,  J''{}_{\g(r_3) \gd(r_3)}(z_3) J_{\a(r_1) \ad(r_1)} (z_1) \rangle \non\\
&=& \frac{1}{k_2} \cI_{\b(r_2) \dot{\l}(r_2)} (x_{2 \bar{1}}) 
\cI_{\l(r_2) \bd(r_2)}(x_{1 \bar{2}}) \cI_{\g(r_3) \dot{\mu}(r_3)} (x_{3 \bar{1}}) \cI_{\mu(r_3) \gd(r_3)} (x_{1 \bar{3}})\non\\
&&\qquad \times \widetilde{H}^{\dot{\l}(r_2) \l(r_2),\, \dot{\mu}(r_3) \m(r_3)}{}_{\a(r_1) \ad(r_1)} (X_1, \Q_1, \bar{\Q}_1)~,
\eea
with $k_2 := (x_{\bar{1} 2} x_{\bar{2} 1})^{r_2+2} (x_{\bar{1} 3} x_{\bar{3} 1})^{r_3+2}$. Making use of the following relations:
\bsubeq
\bea
&&\bar{\cI}^{\ad(r_2) \g(r_2)}(\tilde{x}_{\bar{1} 2}) \cI_{\g(r_2) \bd(r_2)} (x_{2 \bar{3}}) = \bar{\cI}^{\ad(r_2) \g(r_2)} (\bar{X}_1) \cI_{\g(r_2) \bd(r_2)} (x_{1 \bar{3}})~,\\
&&\cI_{\b(r_2) \gd(r_2)} (x_{3 \bar{2}}) \bar{\cI}^{\gd(r_2) \a(r_2)} (\tilde{x}_{\bar{2} 1}) = (-1)^{r_2} \cI_{\b(r_2) \gd(r_2)} (x_{3 \bar{1}}) \bar{\cI}^{\gd(r_2) \a(r_2)} (X_1)~,
\eea
\esubeq
after some manipulations, we obtain
\bsubeq
\bea
\label{e1}
&&\widetilde{H}^{\ad(r_2) \a(r_2), \,\bd(r_3) \b(r_3)}{}_{\g(r_1) \gd(r_1)} (X_1, \Q_1, \bar{\Q}_1) \non\\
&=& (-1)^{r_2}\frac{\big(\l \bar{\l}\big)^{r_2+2}}{\big(x_{\bar{1}3 } x_{\bar{3} 1}\big)^{r_1-r_3}} 
\bar{\cI}^{\ad(r_2) \d(r_2)} (\bar{X}_1) \bar{\cI}^{\dot{\d}(r_2) \a(r_2)} (X_1) \label{tH-H-spinor} \\
&& \times ~\bar{\cI}^{\bd(r_3) \l(r_3)} (\tilde{x}_{\bar{1} 3}) \bar{\cI}^{\dot{\l}(r_3) \b(r_3)} (\tilde{x}_{\bar{3} 1}) \cI_{\d(r_2) \dot{\rho}(r_2)} (x_{1 \bar{3}}) \cI_{\rho(r_2) \dot{\d}(r_2)} (x_{3 \bar{1}}) \non\\
&&\times~ \cI_{\g(r_1) \dot{\s}(r_1)} (x_{1 \bar{3}}) \cI_{\s(r_1) \gd(r_1)} (x_{3 \bar{1}}) H^{\dot{\s}(r_1) \s(r_1), \, \dot{\rho}(r_2) \rho(r_2)}{}_{\l(r_3) \dot{\l}(r_3)} (\fn3)~, ~~~~ \non 
\eea
where 
\bea
\l \bar{\l} = \frac{1}{\big(X_1^2 \bar{X}_1^2 x_{\bar{1} 3}{}^2 x_{\bar{3} 1}{}^2 \big)^{1/2}}~.
\eea
\esubeq
Eq.~\eqref{e1} is quite impractical to work with due to the presence of both two- and three-point functions. It would be desirable to obtain a simpler relation which only involves the three-point building blocks $X, \Q, \bar{\Q}$. Indeed, this procedure can be done via the following steps. First, the analogues of relations \eqref{vect-2pt} allow us to define
\bsubeq \label{vect-2pt-13}
\bea
&&I_{ab}{}^{(13)} \equiv I_{ab}(x_{1 \bar{3}}, x_{\bar{1} 3}) = \hf {\rm tr} (\tilde{\s}_a \, \hat{x}_{1 \bar{3}} \, \tilde{\s}_b \,\hat{x}_{3 \bar{1}})~, \\
&&\bar{I}_{ab}{}^{(31)} \equiv \bar{I}_{ab} (x_{3 \bar{1}}, x_{\bar{3} 1}) = \hf {\rm tr} (\tilde{\s}_a \, \hat{x}_{3 \bar{1}} \, \tilde{\s}_b \,\hat{x}_{1 \bar{3}})~,  \qquad 
I_{ac}{}^{(13)} \bar{I}^{cb\, \,(31)} = \d_{a}{}^{b}~.  
\eea
\esubeq
The operator $I_{ab}{}^{(13)}$ is just an orthogonal matrix with determinant $-1$, that is it satisfies
\bsubeq \label{trf-eta-eps}
\bea
&&I_{ac}{}^{(13)} I_{bd}{}^{(13)} \eta^{cd} = \eta_{ab}~, \\
&&I_{aa'}{}^{(13)}I_{bb'}{}^{(13)}I_{cc'}{}^{(13)}I_{dd'}{}^{(13)}\,\e^{a'b'c'd'} = -\e_{abcd}~.
\eea
\esubeq
With the above definitions, the transformations \eqref{Z13} may be expressed as
\bsubeq \label{trf-X-P}
\bea
I_{ab}{}^{(13)} X_{3}{}^{b} &=& \bar{I}_{ba}{}^{(31)} X_{3}{}^{b} = \l \bar{\l}\, X_{1}{}_{a}^{I}~, \quad
X{}_{a}^I = I_{ab}(\bar{X}, X) X^{b} = - \bigg(\frac{X^2}{\bar{X}^2} \bigg)^{\hf} \bar{X}_{a}~, \\
I_{ab}{}^{(13)} P_{3}{}^{b} &=& \l \bar{\l}\, P_{1}{}_{a}^{I}~, 
\qquad \qquad \qquad \,\,\,\,\,P{}_{a}^I = I_{ab}(\bar{X}, X) P^{b}~,\\
I_{ab}(\bar{X}, X) &=& -\hf {\rm tr} \big( \tilde{\s}_{a}\, \hat{\bar{X}} \,\tilde{\s}_{b} \,\hat{X} \big)~. \label{XbarX-vect}
\eea
\esubeq
In the higher-spin case, from \eqref{vect-2pt-13} and \eqref{XbarX-vect}, one can show that 
\bsubeq
\bea
&&\bar{\cI}^{\ad(k) \b(k)} (\tilde{x}_{\bar{1} 3}) \bar{\cI}^{\bd(k) \a(k)} (\tilde{x}_{\bar{3} 1})\non\\
&=&
\frac{1}{2^k} (\tilde{\s}^{a_1})^{\ad_1 \a_1} \dots (\tilde{\s}^{a_k})^{\ad_k \a_k}(\tilde{\s}^{b_1})^{\bd_1 \b_1} \dots (\tilde{\s}^{b_k})^{\bd_k \b_k} I_{a_1 b_1}^{(13)} \dots I_{a_k b_k}^{(13)}~\non\\
&=&
\frac{1}{2^k} (\tilde{\s}^{a_1})^{\ad_1 \a_1} \dots (\tilde{\s}^{a_k})^{\ad_k \a_k}(\tilde{\s}^{b_1})^{\bd_1 \b_1} \dots (\tilde{\s}^{b_k})^{\bd_k \b_k}\cI_{a(k) b(k)}^{(13)}~,~~~~~~~~~
\eea
as well as
\bea
&&\bar{\cI}^{\bd(k) \a(k)} (\bar{X}) \bar{\cI}^{\ad(k) \b(k)} (X) \non\\
&=&
\bigg(-\hf\bigg)^k (\tilde{\s}^{a_1})^{\ad_1 \a_1} \dots (\tilde{\s}^{a_k})^{\ad_k \a_k}(\tilde{\s}^{b_1})^{\bd_1 \b_1} \dots (\tilde{\s}^{b_k})^{\bd_k \b_k} \cI_{a(k) b(k)}(\bar{X}, X) ~.
\eea
\esubeq
As a result, eq.~\eqref{tH-H-spinor} can be rewritten as follows
\bea \label{tH-23pts-vect}
&&\widetilde{H}^{\ad(r_2) \a(r_2), \,\bd(r_3) \b(r_3)}{}_{\g(r_1) \gd(r_1)} (X_1, \Q_1, \bar{\Q}_1)  \non\\
&=& (-1)^{r_1+r_3}  \frac{\big(\l \bar{\l}\big)^{r_2+2}}{\big(x_{\bar{1}3 } x_{\bar{3} 1}\big)^{r_1-r_3}} 
(\tilde{\s}_{d_1})^{\ad_1 \a_1} \dots (\tilde{\s}_{d_{r_2}})^{\ad_{r_2} \a_{r_2}} (\tilde{\s}^{c_1})^{\bd_1 \b_1} \dots (\tilde{\s}^{c_{r_3}})^{\bd_{r_3} \b_{r_3}} \non\\
&&\quad \times (\s^{a_1})_{\g_1 \gd_1} \dots (\s^{a_{r_1}})_{\g_{r_1} \gd_{r_1}} \cI^{b(r_2) d(r_2)} (\bar{X}_1, X_1) \\
&&\quad \times \,\cI_{a(r_1) a'(r_1)}^{(13)} \, \cI_{b(r_2) b'(r_2)}^{(13)} \,\cI_{c(r_3) c'(r_3)}^{(13)} H^{a'(r_1), b'(r_2), c'(r_3)} (X_3, \Q_3, \bar{\Q}_3)~.\non
\eea
The second step is to compute the last line of \eqref{tH-23pts-vect}, that is:
\bea
&&\cI_{a(r_1) a'(r_1)}^{(13)} \, \cI_{b(r_2) b'(r_2)}^{(13)} \,\cI_{c(r_3) c'(r_3)}^{(13)} H^{a'(r_1), b'(r_2), c'(r_3)} (X_3, \Q_3, \bar{\Q}_3) \non\\
&=& \cI_{a(r_1) a'(r_1)}^{(13)} \, \cI_{b(r_2) b'(r_2)}^{(13)} \,\cI_{c(r_3) c'(r_3)}^{(13)} \non\\
&&\qquad \times \, \Big[F^{a'(r_1), b'(r_2), c'(r_3)} (X_3) + P_{3m} G^{m, \, a'(r_1), b'(r_2), c'(r_3)} (X_3) \Big] \\
&=& \cI_{a(r_1) a'(r_1)}^{(13)} \, \cI_{b(r_2) b'(r_2)}^{(13)} \,\cI_{c(r_3) c'(r_3)}^{(13)} F^{a'(r_1), b'(r_2), c'(r_3)}(X_3)\non\\
&&+ (\l \bar{\l}) \,\cI_{a(r_1) a'(r_1)}^{(13)} \, \cI_{b(r_2) b'(r_2)}^{(13)} \,\cI_{c(r_3) c'(r_3)}^{(13)} \bar{I}_{m m'}^{\,\,(31)} \, P_1{}^{I m'} G^{m, \, a'(r_1), b'(r_2), c'(r_3)} (X_3)~, \non
\eea
where, in the last equality above we have expressed $P_3$ in terms of $P_1{}^I$. Next, we note that $F(X_3)$ and $G(X_3)$ (indices suppressed) can 
only be constructed out of $X_{3}^a, \eta_{ab}$ and $\e_{abcd}$. Though at this stage we do not know yet the explicit solution/ tensorial structure, 
we can always express the solution as $F(X_3) = F_1(X_3) + F_2(X_3)$, where all the dependence on $\e_{abcd}$ resides in $F_2$. 
The reason for such a splitting is due to the transformation properties \eqref{trf-eta-eps} and \eqref{trf-X-P}. These, along with the homogeneity 
property of $F(X_3)$, tell us that $F_1 (X_3)$ transforms as a tensor, while $F_2 (X_3)$ transforms as a pseudotensor under the actions of the two-point 
operators $\cI_{a(k) b(k)}^{(13)}$. More precisely:
\bea
&&\cI_{a(r_1) a'(r_1)}^{(13)} \, \cI_{b(r_2) b'(r_2)}^{(13)} \,\cI_{c(r_3) c'(r_3)}^{(13)} F^{a'(r_1), b'(r_2), c'(r_3)}(X_3)\non\\
&=& (\l \bar{\l})^{r_3-(r_1+r_2+2)} \Big[ (F_1)_{a(r_1),\, b(r_2),\, c(r_3)} (X_1{}^I) - (F_2)_{a(r_1),\, b(r_2),\, c(r_3)} (X_1{}^I) \Big]~. \label{x-F}
\eea
The same argument also applies to $G(X_3)$. We find that
\bea
&&(\l \bar{\l}) \,\cI_{a(r_1) a'(r_1)}^{(13)} \, \cI_{b(r_2) b'(r_2)}^{(13)} \,\cI_{c(r_3) c'(r_3)}^{(13)} \bar{I}_{m m'}^{\,\,(31)} \, P_1{}^{I m'} G^{m, \, a'(r_1), b'(r_2), c'(r_3)} (X_3) \non\\
&=& (\l \bar{\l})^{r_3-(r_1+r_2+2)} P_1{}^{I m} \Big[ (G_1)_{m, \, a(r_1),\, b(r_2),\, c(r_3)} (X_1{}^I) - (G_2)_{m, \, a(r_1),\, b(r_2),\, c(r_3)} (X_1{}^I) \Big]~.~~~~~~ \label{x-G}
\eea
Eq.\,\eqref{tH-23pts-vect} then turns into
\bea \label{tH-vect-X}
&&\widetilde{H}^{\ad(r_2) \a(r_2), \,\bd(r_3) \b(r_3)}{}_{\g(r_1) \gd(r_1)} (X_1, \Q_1, \bar{\Q}_1)  \non\\
&=& \bigg(  \frac{-1}{X_1 \bar{X}_1} \bigg)^{r_3-r_1} 
(\tilde{\s}_{d_1})^{\ad_1 \a_1} \dots (\tilde{\s}_{d_{r_2}})^{\ad_{r_2} \a_{r_2}} (\tilde{\s}^{c_1})^{\bd_1 \b_1} \dots (\tilde{\s}^{c_{r_3}})^{\bd_{r_3} \b_{r_3}} \non\\
&&\quad \times (\s^{a_1})_{\g_1 \gd_1} \dots (\s^{a_{r_1}})_{\g_{r_1} \gd_{r_1}} \cI^{b(r_2) d(r_2)} (\bar{X}_1, X_1) \non\\
&&\quad \times \bigg\{ (F_1)_{a(r_1),\, b(r_2),\, c(r_3)} (X_1{}^I) - (F_2)_{a(r_1),\, b(r_2),\, c(r_3)} (X_1{}^I) \\
&& \qquad + P_1{}^{I m} \Big[(G_1)_{m, \, a(r_1),\, b(r_2),\, c(r_3)} (X_1{}^I) - (G_2)_{m, \, a(r_1),\, b(r_2),\, c(r_3)} (X_1{}^I) \Big]
\bigg\}~.\non
\eea
The last step is thus to express $\bar{X}$ in terms of $X$ and $P$, with the help of identities \eqref{XbarX} and \eqref{trf-X-P}. For instance, since $F_{a(r_1),\, b(r_2),\, c(r_3)} (X)$ is a homogeneous function and has a scale dimension $d$, it is possible to Taylor's expand the function and obtain
\bea
&&F_{a(r_1),\, b(r_2),\, c(r_3)} (X^I) = \bigg( -\frac{X}{\bar{X}} \bigg)^{d}F_{a(r_1),\, b(r_2),\, c(r_3)} (\bar{X}) \non\\
&=& \bigg( -\frac{X}{\bar{X}} \bigg)^{d} \bigg[ F_{a(r_1),\, b(r_2),\, c(r_3)}(X) + \ri P^m \pa_{m} F_{a(r_1),\, b(r_2),\, c(r_3)}(X) \non\\
&& \qquad \qquad \quad- \frac{1}{8} P^2 \Box F_{a(r_1),\, b(r_2),\, c(r_3)}(X) \bigg] \non\\
&=&(-1)^{d}  \bigg\{ F_{a(r_1),\, b(r_2),\, c(r_3)}(X) \non\\
&& \qquad \quad + \ri P^m \Big[ \pa_m  F_{a(r_1),\, b(r_2),\, c(r_3)}(X) - d\, \frac{X_m}{X^2}  F_{a(r_1),\, b(r_2),\, c(r_3)}(X)\Big] \non\\
&& \qquad \quad -\frac{1}{8} P^2 \Big[ \Box F_{a(r_1),\, b(r_2),\, c(r_3)}(X) - d(d+2) \frac{1}{X^2} F_{a(r_1),\, b(r_2),\, c(r_3)}(X) \Big] \bigg\}~.
\eea
In addition, the final result can be presented in a more compact way once we contract all indices with the auxiliary commuting spinors. We define
\bea
\widetilde{H}(X,P;\, u, \bar{u}, v, \bar{v}, w, \bar{w}) &=&  
\mathbf{U}^{\ad(r_2)\a(r_2) } \mathbf{V}^{\bd(r_3) \b(r_3) } \mathbf{W}^{\gd(r_1) \g(r_1) } \non\\
&&\quad \times~ \widetilde{H}_{\a(r_2) \ad(r_2) ,\,\b(r_3) \bd(r_3), \, \g(r_1) \gd(r_1)} (X_1, \Q_1, \bar{\Q}_1)~.
\eea
After some lengthy computations and recalling that $\widetilde{H}(X,P;\, u, \bar{u}, v, \bar{v}, w, \bar{w})$ has scale dimension $r_1-(r_2+r_3+2)$, one can present it in the form
\bea
\widetilde{H}(X,P;\, u, \bar{u}, v, \bar{v}, w, \bar{w}) &=&  
\widetilde{F}({X};\, u, \bar{u}, v, \bar{v}, w, \bar{w}) -\hf P^{\ad \a}\widetilde{G}_{\a \ad}({X};\, u, \bar{u}, v, \bar{v}, w, \bar{w})\non\\
&&~+ P^2 \widetilde{K}({X};\, u, \bar{u}, v, \bar{v}, w, \bar{w})~,~~~~~~~~~
\eea
where we have defined
\bsubeq
\bea
&&\widetilde{F}({X};\, u, \bar{u}, v, \bar{v}, w, \bar{w}) = \frac{(-1)^{r_2}}{r_2! r_2!} \frac{1}{X^{2(r_3+r_2-r_1)}}\big(u {X} \pa_{(\bar{s})}\big)^{r_2}  \big(\bar{u} {X} \pa_{({s})}\big)^{r_2} \non\\
&&\qquad \qquad \qquad \qquad \qquad \quad \times~ \Big[F_1({X}; w, \bar{w}, s, \bar{s}, v, \bar{v}) - F_2 ({X}; w, \bar{w}, s, \bar{s}, v, \bar{v})\Big]~,~~~~~~  \label{tildeF}\\
&&\widetilde{G}_{\a \ad}({X};\, u, \bar{u}, v, \bar{v}, w, \bar{w}) = \frac{(-1)^{r_2}}{r_2! r_2!} \frac{1}{X^{2(r_3+r_2-r_1)}} \big(u {X} \pa_{(\bar{s})}\big)^{r_2} \non\\
&&\qquad \times~ \bigg\{ 2 \ri(r_1-r_3+1)  \big(\bar{u} {X} \pa_{({s})}\big)^{r_2} \frac{{X}_{\a \ad}}{X^2} \Big[F_1({X}; w, \bar{w}, s, \bar{s}, v, \bar{v}) - {F}_2 ({X}; w, \bar{w}, s, \bar{s}, v, \bar{v}) \Big] \non\\
&&\qquad \quad - 2\ri r_2 \big(\bar{u} {X} \pa_{({s})}\big)^{r_2-1} \bar{u}_{\ad} \frac{\pa}{\pa s^{\a}}\Big[F_1(X; w, \bar{w}, s, \bar{s}, v, \bar{v}) - F_2 (X; w, \bar{w}, s, \bar{s}, v, \bar{v}) \Big] \non\\
&& \qquad \quad + \ri \big(\bar{u} X \pa_{({s})}\big)^{r_2} \pa_{\a \ad} \Big[F_1(X; w, \bar{w}, s, \bar{s}, v, \bar{v}) - F_2 (X; w, \bar{w}, s, \bar{s}, v, \bar{v}) \Big] \non\\
&& \qquad \quad - \big(\bar{u} X \pa_{({s})}\big)^{r_2} \frac{X_{\a\bd} X_{\b \ad}}{X^2} 
\Big[G_1{}^{\bd \b}(X; w, \bar{w}, s, \bar{s}, v, \bar{v}) - G_2{}^{\bd \b} (X; w, \bar{w}, s, \bar{s}, v, \bar{v}) \Big]
\bigg\}~,~~~~~~~~~~~ \label{tildeG}\\
&&\widetilde{K}({X};\, u, \bar{u}, v, \bar{v}, w, \bar{w}) =
\frac{(-1)^{r_2}}{r_2! r_2!} \frac{1}{X^{2(r_3+r_2-r_1)}} \big(u {X} \pa_{(\bar{s})}\big)^{r_2} \non\\
&&\qquad \times~\bigg\{ -\frac{1}{8}\big(\bar{u} {X} \pa_{({s})}\big)^{r_2} \Big[ \Box F_1(X; w, \bar{w}, s, \bar{s}, v, \bar{v}) - \Box F_2(X; w, \bar{w}, s, \bar{s}, v, \bar{v}) \non\\
&&\qquad \qquad \qquad \qquad \quad~- \ri \pa^{\ad \a} G_{1 \a \ad}(X; w, \bar{w}, s, \bar{s}, v, \bar{v})+ \ri \pa^{\ad \a} G_{2 \a \ad}(X; w, \bar{w}, s, \bar{s}, v, \bar{v}) \Big] \non\\
&&\qquad \quad-\frac{\ri r_2}{4} \big(\bar{u} {X} \pa_{({s})}\big)^{r_2-1} \bar{u}^{\ad} \ve^{\a \b} \frac{\pa}{\pa s^{\b}}
\Big[ G_{1\, \a \ad}(X; w, \bar{w}, s, \bar{s}, v, \bar{v}) - G_{2\, \a \ad}(X; w, \bar{w}, s, \bar{s}, v, \bar{v}) \non\\
&&\qquad \qquad \qquad \qquad \quad~- \ri \pa_{\a \ad} F_1(X; w, \bar{w}, s, \bar{s}, v, \bar{v})+ \ri \pa_{\a \ad} F_2(X; w, \bar{w}, s, \bar{s}, v, \bar{v}) \Big] \bigg\}~.
\eea
\esubeq
Here we have made use of the notation 
\bea
\big(uX \pa_{(\bar{s})} \big) = u_{\a}X^{\ad \a} \frac{\pa}{\pa \bar{s}^{\ad}}~, 
\qquad \big( \bar{u} X \pa_{({s})} \big) = \bar{u}_{\ad}X^{\ad \a} \frac{\pa}{\pa {s}^{\a}}~.
\eea
It can be checked that $\widetilde{K}({X};\, u, \bar{u}, v, \bar{v}, w, \bar{w})$ vanishes due to conservation conditions on the first two points. 
This means that 
\be
\widetilde{H}(X,P;\, u, \bar{u}, v, \bar{v}, w, \bar{w}) =
\widetilde{F}({X};\, u, \bar{u}, v, \bar{v}, w, \bar{w}) -\hf P^{\ad \a}\widetilde{G}_{\a \ad}({X};\, u, \bar{u}, v, \bar{v}, w, \bar{w})
\label{e2}
\ee
with $\widetilde{F}$ and $\widetilde{G}$ given by eqs.~\eqref{tildeF} and~\eqref{tildeG}.

Having derived the full expression for $\widetilde{H}(X,P;\, u, \bar{u}, v, \bar{v}, w, \bar{w})$ which appears in the correlator
\bea
&&\langle
J'{}_{\b(r_2) \bd(r_2)}(z_2)\,  J''{}_{\g(r_3) \gd(r_3)}(z_3) J_{\a(r_1) \ad(r_1)} (z_1) \rangle \non\\
&=& \frac{1}{k_2} \cI_{\b(r_2) \dot{\l}(r_2)} (x_{2 \bar{1}}) 
\cI_{\l(r_2) \bd(r_2)}(x_{1 \bar{2}}) \cI_{\g(r_3) \dot{\mu}(r_3)} (x_{3 \bar{1}}) \cI_{\mu(r_3) \gd(r_3)} (x_{1 \bar{3}})\non\\
&&\qquad \times \widetilde{H}^{\dot{\l}(r_2) \l(r_2),\, \dot{\mu}(r_3) \m(r_3)}{}_{\a(r_1) \ad(r_1)} (X_1, \Q_1, \bar{\Q}_1)~,
\eea
we can then impose conservation conditions on $J''{}_{\g(r_3) \gd(r_3)}$ as if it were located at the ``second point''. The required  constraints take the form
\bsubeq
\bea
&&\ve^{\a \b} \frac{\pa}{\pa v^{\b}} \widetilde{G}_{\a \ad} (X;\, u, \bar{u}, v, \bar{v}, w, \bar{w})= 0~, \\
&&\pa^{\bd \a} \frac{\pa}{\pa \bar{v}^{\bd}} \widetilde{G}_{\a \ad} (X;\, u, \bar{u}, v, \bar{v}, w, \bar{w}) = 0~,\\
&&\ve^{\ad \bd} \frac{\pa}{\pa \bar{v}^{\bd}} \Big[ \widetilde{G}_{\a \ad} (X;\, u, \bar{u}, v, \bar{v}, w, \bar{w}) - \ri \pa_{\a \ad} \widetilde{F} (X;\, u, \bar{u}, v, \bar{v}, w, \bar{w}) \Big] = 0~,
\eea
\esubeq
which also give rise to the consistency condition
\bea
\Box \widetilde{F}(X;\, u, \bar{u}, v, \bar{v}, w, \bar{w}) - \ri \pa^{\ad \a} \widetilde{G}_{\a \ad} (X;\, u, \bar{u}, v, \bar{v}, w, \bar{w})= 0~. 
\eea


\subsection{Summary of the constraints}


To conclude this section, here we list out all the constraints imposed on our correlation function, using the generating function formalism. The first main point is that there are different representations, which correspond to inequivalent ansatz for how to write the same correlator:
\bea 
&&\langle
J_{\a(r_1) \ad(r_1)} (z_1) \, J'{}_{\b(r_2) \bd(r_2)}(z_2)\,  J''{}_{\g(r_3) \gd(r_3)}(z_3)
\rangle \non\\
&=& \frac{1}{k_1} \cI_{\a(r_1) \dot{\mu}(r_1)} (x_{1 \bar{3}}) \cI_{{\mu}(r_1) \ad(r_1)} (x_{3 \bar{1}}) \cI_{\b(r_2) \dot{\l}(r_2)} (x_{2 \bar{3}}) \cI_{{\l}(r_2) \bd(r_2)} (x_{3 \bar{2}})\non\\
&& \qquad \times H^{\dot{\mu}(r_1) \mu(r_1), \, \dot{\l}(r_2) \l(r_2)}{}_{\g(r_3) \dot{\g}(r_3)} (X_3, P_3) \non\\
&=& \frac{1}{k_2} \cI_{\b(r_2) \dot{\l}(r_2)} (x_{2 \bar{1}}) 
\cI_{\l(r_2) \bd(r_2)}(x_{1 \bar{2}}) \cI_{\g(r_3) \dot{\mu}(r_3)} (x_{3 \bar{1}}) \cI_{\mu(r_3) \gd(r_3)} (x_{1 \bar{3}})\non\\
&&\qquad \times \widetilde{H}^{\dot{\l}(r_2) \l(r_2),\, \dot{\mu}(r_3) \m(r_3)}{}_{\a(r_1) \ad(r_1)} (X_1, P_1)~, \label{general-ansatz}
\eea 
where $k_1 := (x_{\bar{1} 3} x_{\bar{3} 1})^{r_1+2} (x_{\bar{2} 3} x_{\bar{3} 2})^{r_2+2}$ and $k_2 := (x_{\bar{1} 2} x_{\bar{2} 1})^{r_2+2} (x_{\bar{1} 3} x_{\bar{3} 1})^{r_3+2}$.
The general solution for the tensor $H$ can always be presented in the form
\be
\begin{aligned}
&H_{\a(r_1) \ad(r_1), \, \b(r_2) \bd(r_2), \, \g(r_3) \gd(r_3)} (\fxq) \\
&=F_{\a(r_1) \ad(r_1), \, \b(r_2) \bd(r_2), \, \g(r_3) \gd(r_3)} ({X})\\
&\quad -\hf P^{\dot{\d} \d} G_{\d \dot{\d},\,\a(r_1) \ad(r_1), \, \b(r_2) \bd(r_2), \, \g(r_3) \gd(r_3)} ({X})~.~~~~~~~
\end{aligned}
\ee
We then associate to $H_{\a(r_1) \ad(r_1), \, \b(r_2) \bd(r_2), \, \g(r_3) \gd(r_3)} (X, P)$ the polynomial 
\bea
&& H(X,P; u, \bar{u}, v, \bar{v}, w, \bar{w}) \non\\
&=& \mathbf{U}^{\ad(r_1) \a(r_1) } \mathbf{V}^{\bd(r_2) \b(r_2) } \mathbf{W}^{\gd(r_3) \g(r_3) } H_{\a(r_1) \ad(r_1), \, \b(r_2) \bd(r_2), \, \g(r_3) \gd(r_3)} (X, P) \non\\
&=& F (X;\, u, \bar{u}, v, \bar{v}, w, \bar{w}) -\hf P^{\ad \a} G_{\a \ad} (X;\, u, \bar{u}, v, \bar{v}, w, \bar{w})~,
\eea
and to $\widetilde{H}_{\a(r_2) \ad(r_2) ,\,\b(r_3) \bd(r_3), \, \g(r_1) \gd(r_1)} (X, P)$ the polynomial
\bea
&&\widetilde{H}(X,P;\, u, \bar{u}, v, \bar{v}, w, \bar{w}) \non\\ 
&=& \mathbf{U}^{\ad(r_2)\a(r_2) } \mathbf{V}^{\bd(r_3) \b(r_3) } \mathbf{W}^{\gd(r_1) \g(r_1) } \widetilde{H}_{\a(r_2) \ad(r_2) ,\,\b(r_3) \bd(r_3), \, \g(r_1) \gd(r_1)} (X,P) \non\\
&=& \widetilde{F} (X;\, u, \bar{u}, v, \bar{v}, w, \bar{w}) -\hf P^{\ad \a} \widetilde{G}_{\a \ad} (X;\, u, \bar{u}, v, \bar{v}, w, \bar{w})~.
\eea
Our formalism essentially allows us to express $\widetilde{F}$ and $\widetilde{G}_{\a \ad}$ in terms of only $F$, $G_{\a \ad}$ and their vector derivatives (see eqs.~\eqref{tildeF} and \eqref{tildeG} for the explicit relations). This is useful since the polynomials $F({X}; u, \bar{u}, v, \bar{v}, w, \bar{w})$ and $G_{\a \ad}({X}; u, \bar{u}, v, \bar{v}, w, \bar{w})$ are functions of a \textit{single, bosonic} covariant building block $X$ (and the auxiliary commuting spinors).
When written in terms of  $F$, $G_{\a \ad}$, $\widetilde{F}$ and $\widetilde{G}_{\a \ad}$, the constraints due to conservation laws at all points take a simple form. More precisely, all the required constraints are:
\begin{itemize}
\item[\textbf{(1)}] \textbf{Homogeneity:}
\bsubeq
\bea
&&F ( \L \bar{\L}X;\, u, \bar{u}, v, \bar{v}, w, \bar{w}) = (\L \bar{\L})^{r_3-(r_1+r_2+2)} F (X;\, u, \bar{u}, v, \bar{v}, w, \bar{w})~, \\
&&G_{\a \ad} (\L \bar{\L} X;\, u, \bar{u}, v, \bar{v}, w, \bar{w}) = (\L \bar{\L})^{r_3-(r_1+r_2+3)} G_{\a \ad} (X;\, u, \bar{u}, v, \bar{v}, w, \bar{w})~. ~~~~~~
\eea
\esubeq
\item[\textbf{(2)}] \textbf{Conservation conditions at all points:}
\bsubeq
\bea
&&\ve^{\ad \bd} \frac{\pa}{\pa \bar{u}^{\bd}} G_{\a \ad}({X}; u, \bar{u}, v, \bar{v}, w, \bar{w}) = 0~, \label{conserv-1}\\
&&\pa^{\ad \b} \frac{\pa}{\pa u^{\b}} G_{\a \ad} (X; u, \bar{u}, v, \bar{v}, w, \bar{w}) = 0~, \\
&&\ve^{\a \b} \frac{\pa}{\pa u^{\b}} \Big[ G_{\a \ad}(X; u, \bar{u}, v, \bar{v}, w, \bar{w})- \ri  \pa_{\a \ad} F(X; u, \bar{u}, v, \bar{v}, w, \bar{w}) \Big] = 0~,\\
&&\pa^{\bd \a} \frac{\pa}{\pa \bar{v}^{\bd}} G_{\a \ad} (X; u, \bar{u}, v, \bar{v}, w, \bar{w}) = 0~, \\
&&\ve^{\ad \bd} \frac{\pa}{\pa \bar{v}^{\bd}} \Big[ G_{\a \ad}(X; u, \bar{u}, v, \bar{v}, w, \bar{w})- \ri  \pa_{\a \ad} F(X; u, \bar{u}, v, \bar{v}, w, \bar{w}) \Big] = 0~,\\
&&\ve^{\a \b} \frac{\pa}{\pa {v}^{\b}} G_{\a \ad}({X}; u, \bar{u}, v, \bar{v}, w, \bar{w}) = 0~, \label{conserv-2-end}\\
&&\ve^{\a \b} \frac{\pa}{\pa v^{\b}} \widetilde{G}_{\a \ad} (X;\, u, \bar{u}, v, \bar{v}, w, \bar{w})= 0~, \label{conserv-3}\\
&&\pa^{\bd \a} \frac{\pa}{\pa \bar{v}^{\bd}} \widetilde{G}_{\a \ad} (X;\, u, \bar{u}, v, \bar{v}, w, \bar{w}) = 0~,\\
&&\ve^{\ad \bd} \frac{\pa}{\pa \bar{v}^{\bd}} \Big[ \widetilde{G}_{\a \ad} (X;\, u, \bar{u}, v, \bar{v}, w, \bar{w}) - \ri \pa_{\a \ad} \widetilde{F} (X;\, u, \bar{u}, v, \bar{v}, w, \bar{w}) \Big] = 0~. \label{conserv-3-end}
\eea
\esubeq
The above differential constraints lead to consistency conditions
\bsubeq
\bea
&&\Box {F}(X;\, u, \bar{u}, v, \bar{v}, w, \bar{w}) - \ri \pa^{\ad \a}{G}_{\a \ad} (X;\, u, \bar{u}, v, \bar{v}, w, \bar{w})= 0~, \\
&&\Box \widetilde{F}(X;\, u, \bar{u}, v, \bar{v}, w, \bar{w}) - \ri \pa^{\ad \a} \widetilde{G}_{\a \ad} (X;\, u, \bar{u}, v, \bar{v}, w, \bar{w})= 0~. 
\eea
\esubeq
\item[\textbf{(3)}] \textbf{Reality:} $a_i$ are real and
\bea
G_{\a \ad}({X}; u, \bar{u}, v, \bar{v}, w, \bar{w};\,{b}_i) &=&
G_{\a \ad} (X; u, \bar{u}, v, \bar{v}, w, \bar{w}; \,\bar{b}_i) \non\\ 
&+& \ri \pa_{\a \ad} F(X; u, \bar{u}, v, \bar{v}, w, \bar{w}; {a}_i)~.
\eea
Here $a_i$ and $b_i$ are constant parameters of the linearly independent tensor structures in $F(X; u, \bar{u}, v, \bar{v}, w, \bar{w})$ and $G_{\a \ad}(X; u, \bar{u}, v, \bar{v}, w, \bar{w})$, respectively.
\end{itemize}

In the remaining sections we will apply our formalism to specific examples. We will discussed mixed correlation functions 
of $J_{\a(r) \ad(r)}$ with the supercurrent and the flavour current multiplets. 


\section{Correlator \texorpdfstring{$\la J_{\a(r) \ad(r)} (z_1) J_{\b \bd} (z_2) J_{\g \gd} (z_3) \ra$}{<r11>}}
\label{section4}


In this section we compute a mixed three-point function containing the higher-spin supercurrent $J_{\a(r) \ad(r)}$ and the supercurrent $J_{\a \ad}$. In accordance with \eqref{general-ansatz}, the ansatz for this correlator has the form
\bea
\langle
J_{\a(r) \ad(r)} (z_1) \, J_{\b \bd}(z_2)\,  J_{\g \gd}(z_3)
\rangle 
&=& 
\frac{1}{(x_{\bar{1} 3} x_{\bar{3} 1})^{r+2} (x_{\bar{2} 3} x_{\bar{3} 2})^{3}} \cI_{\a(r) \dot{\mu}(r)} (x_{1 \bar{3}}) \cI_{{\mu}(r)\ad(r)} (x_{3 \bar{1}}) \non\\
&&\quad \times~ \cI_{\b \dot{\l}} (x_{2 \bar{3}}) \cI_{{\l} \bd} (x_{3 \bar{2}}) H^{\dot{\mu}(r) \mu(r), \, \dot{\l} \l}{}_{\g \dot{\g}} (X_3, P_3)~,
\eea
where $H_{\a(r) \dot{\a}(r), \,\b \dot{\b}, \,\g \gd}$ has dimension $-(r+2)$. The case $r=1$ corresponds to the three-point function of the ordinary $\cN=1$ supercurrent, which was first derived by Osborn \cite{OsbornN1}. In subsection \ref{subsect4.1}, we will rederive this result using the procedure outlined in section \ref{section3}, thus providing a nice check of our formalism. It should be noted that the explicit solution for $H_{\a \ad, \b \bd, \g \gd}$ constructed in \cite{OsbornN1} is expressed in terms of the three-point building blocks $X$, $\bar{X}$ and $P$. Hence, in order to make contact with our result, one needs to replace $\bar{X}$ everywhere using the relation $\bar{X} = X+ \ri P$. As the next step, it is natural to consider the case for arbitrary $r$. In constructing the tensorial structure for $G_{\a \ad}$, it is necessary to treat the $r=2$ and $r > 2 $ cases separately.

\subsection{Example: Correlator \texorpdfstring{$\la J_{\a \ad} (z_1) J_{\b \bd} (z_2) J_{\g \gd} (z_3) \ra$}{<j1 j1 j1>}} \label{subsect4.1}
In this particular case, the solution for $H_{\a \dot{\a}, \b \dot{\b}, \g \gd}$ has a scale dimension $-3$. We then construct a generating function defined as follows:
\bea
H(X,P; \, u, \bar{u}, v, \bar{v}, w, \bar{w})
&=& \mathbf{U}^{\ad \a } \mathbf{V}^{\bd \b } \mathbf{W}^{\gd \g } H_{\a\ad, \, \b \bd, \, \g \gd} (X, P) \non\\
&=& F (X;\, u, \bar{u}, v, \bar{v}, w, \bar{w}) -\hf P^{\ad \a} G_{\a \ad} (X;\, u, \bar{u}, v, \bar{v}, w, \bar{w})~.~~~~~~
\eea
The general expansion for the polynomial $F (X;\, u, \bar{u}, v, \bar{v}, w, \bar{w})$ can be written as
\bea
F (X;\, u, \bar{u}, v, \bar{v}, w, \bar{w}) = F_1 (X;\, u, \bar{u}, v, \bar{v}, w, \bar{w}) +  F_2 (X;\, u, \bar{u}, v, \bar{v}, w, \bar{w})~,
\eea
where
\bea
F_1 (X;\, u, \bar{u}, v, \bar{v}, w, \bar{w}) = \frac{1}{X^3}\cF_1(\hat{X}; u, \bar{u}, v, \bar{v}, w, \bar{w} )~.
\eea
Here the polynomial $\cF_1$ is homogeneous of degree 0 in $X$ and degree 1 in $u, \bar{u}, v, \bar{v}, w, \bar{w}$. It is not difficult to construct all possible polynomial structures for $\cF_1$ in terms of our basis structures described in \eqref{basis-F}. We find that
\bea
&&\cF_1(\hat{X}; u, \bar{u}, v, \bar{v}, w, \bar{w} )\non\\
&=& a_1 (U \hat{X}) (V \hat{X}) (W \hat{X}) + a_2 (V W) (U \hat{X}) + a_3 (U V) (W \hat{X}) + a_4 (U W) (V \hat{X}) \non\\
&=& -\frac{a_1}{8} \hat{X}_{(u,\bar{u})} \hat{X}_{(v,\bar{v})} \hat{X}_{(w,\bar{w})} + \frac{a_2}{4} \hat{X}_{(u,\bar{u})}(v \cdot w)(\bar{v}\cdot \bar{w}) +\frac{a_3}{4} \hat{X}_{(w,\bar{w})}(u\cdot v)(\bar{u}\cdot \bar{v}) \non\\
&&+ \frac{a_4}{4} \hat{X}_{(v,\bar{v})}(u\cdot w)(\bar{u} \cdot \bar{w})~.
\eea
By a similar argument, we have that
\bea
F_2 (X;\, u, \bar{u}, v, \bar{v}, w, \bar{w}) &=& \frac{1}{X^3}\cF_2(\hat{X}; u, \bar{u}, v, \bar{v}, w, \bar{w} ).
\eea
It is clear that there is only one possible structure
\bea
&&\cF_2(\hat{X}; u, \bar{u}, v, \bar{v}, w, \bar{w} ) = a_5 \,\cJ \non\\
&=& \frac{\ri}{4} a_5 \big[ \hat{X}_{(u, \bar{w})} (v \cdot w) (\bar{u} \cdot \bar{v}) - \hat{X}_{(w, \bar{u})} (u \cdot v) (\bar{v} \cdot \bar{w})
\big]~.
\eea

We shall now find the general expansion for the vector $G_{\a \ad}(X;\, u, \bar{u}, v, \bar{v}, w, \bar{w})$:
\bea
G_{\a \ad}(X;\, u, \bar{u}, v, \bar{v}, w, \bar{w}) = \frac{1}{X^5} \cG_{\a \ad}(\hat{X};\, u, \bar{u}, v, \bar{v}, w, \bar{w})~.
\eea
We can now decompose $\cG$ further using the structures defined in \eqref{basis-G}:
\bea
\cG_{\a \ad}(\hat{X};\, u, \bar{u}, v, \bar{v}, w, \bar{w}) = \sum_{I=1}^{8} \cK_{I, \a \ad} \bm{\cG}_{I}(\hat{X};\, u, \bar{u}, v, \bar{v}, w, \bar{w})~.
\eea
Here it should be noted that $\bm{\cG}_{I}$ now are polynomials which are homogeneous of degree 0 in $X$, degree 2 in $U$, and degree 1 in both $V$ and $W$. We find that:
\bsubeq
\bea
&&\text{$\cK_1$ structures:}\non\\
&&\qquad \quad \cK_{1, \a \ad} \bm{\cG}_{1}(\hat{X};\, u, \bar{u}, v, \bar{v}, w, \bar{w}) = u_{\a} \bar{u}_{\ad} \Big[ b_1 (V \hat{X}) (W \hat{X}) + b_2 (VW) \Big]~,\\
&&\text{$\cK_2$ structures:}\non\\
&&\qquad \quad \cK_{2, \a \ad} \bm{\cG}_{2}(\hat{X};\, u, \bar{u}, v, \bar{v}, w, \bar{w}) = \hat{X}_{\a \ad} \Big[ b_3 (U \hat{X}) (V \hat{X}) (W \hat{X}) + b_4 (U \hat{X}) (VW)  \non\\
&&\qquad \qquad \qquad \qquad + b_5 (V \hat{X}) (UW) + b_6 (W \hat{X}) (UV) + b_7 \cJ \Big]~,\\
&&\text{$\cK_3$ structures:}\non\\
&&\qquad \quad \cK_{3, \a \ad} \bm{\cG}_{3}(\hat{X};\, u, \bar{u}, v, \bar{v}, w, \bar{w}) = v_{\a} \bar{v}_{\ad} \Big[ b_8 (UW)  + b_9 (U \hat{X}) (W \hat{X}) \Big]~,\\
&&\text{$\cK_4$ structures:}\non\\
&&\qquad \quad \cK_{4, \a \ad} \bm{\cG}_{4}(\hat{X};\, u, \bar{u}, v, \bar{v}, w, \bar{w}) = w_{\a} \bar{w}_{\ad} \Big[ b_{10} (UV)  + b_{11} (U \hat{X}) (V \hat{X}) \Big]~,\\
&&\text{$\cK_5$ structure:}\non\\
&&\qquad \quad \cK_{5, \a \ad} \bm{\cG}_{5}(\hat{X};\, u, \bar{u}, v, \bar{v}, w, \bar{w}) = \cZ_{1\, \a \ad} \big[ b_{12} (W \hat{X})\big]~,\\
&&\text{$\cK_6$ structure:}\non\\
&&\qquad \quad \cK_{6, \a \ad} \bm{\cG}_{6}(\hat{X};\, u, \bar{u}, v, \bar{v}, w, \bar{w}) = \cZ_{2\, \a \ad} \big[ b_{13} (V \hat{X})\big]~,\\
&&\text{$\cK_7$ structure:}\non\\
&&\qquad \quad \cK_{7, \a \ad} \bm{\cG}_{7}(\hat{X};\, u, \bar{u}, v, \bar{v}, w, \bar{w}) =\cZ_{3\, \a \ad} \big[ b_{14} (U \hat{X})\big]~,\\ 
&&\text{$\cK_8$ structure:}\non\\
&&\qquad \quad \cK_{8, \a \ad} \bm{\cG}_{8}(\hat{X};\, u, \bar{u}, v, \bar{v}, w, \bar{w}) = b_{15}\, \cZ_{4\, \a \ad}~.
\eea
\esubeq
It is worth noting that not all of the $\cK$-structures are linearly independent. In particular, we see that
\bea
\hat{X}_{\a \ad} \cJ + (W \hat{X}) \cZ_{1\, \a \ad} - (V \hat{X})\cZ_{2\, \a \ad} + (U \hat{X}) \cZ_{3\, \a \ad} - \cZ_{4\, \a \ad} = 0~. \label{lin-dep-1}
\eea  
Thus, we choose to construct a linearly independent basis of polynomial structures for the vector $G_{\a \ad}(X;\, u, \bar{u}, v, \bar{v}, w, \bar{w})$ by removing the $b_{7}$ structure.
As a result, we can express
\bsubeq
\bea
&&G_{\a \ad}(X;\, u, \bar{u}, v, \bar{v}, w, \bar{w}) = G_{1\,\a \ad}(X;\, u, \bar{u}, v, \bar{v}, w, \bar{w})+ G_{2\, \a \ad}(X;\, u, \bar{u}, v, \bar{v}, w, \bar{w})~,~~~~~~~\\
&&G_{1\,\a \ad}(X;\, u, \bar{u}, v, \bar{v}, w, \bar{w}) = \frac{1}{X^4} \sum_{I=1}^{4} \cK_{I, \a \ad} \bm{\cG}_{I}(\hat{X};\, u, \bar{u}, v, \bar{v}, w, \bar{w})~,\\
&&G_{2\,\a \ad}(X;\, u, \bar{u}, v, \bar{v}, w, \bar{w}) = \frac{1}{X^4} \sum_{I=5}^{8} \cK_{I, \a \ad} \bm{\cG}_{I}(\hat{X};\, u, \bar{u}, v, \bar{v}, w, \bar{w})~.
\eea
\esubeq
We also note that all $a_i$ and $b_i$ are still assumed to be complex coefficients. 

The next step is to impose differential constraints arising from the conservation on the first two points, eqs.~\eqref{conserv-1}--\eqref{conserv-2-end}. This can be computed quickly using \textit{Mathematica}. We find that only 4 independent complex coefficients remain, which we choose to be $a_1, a_2, a_5, b_1$. Reality condition implies that $a_1, a_2, a_5$ are all real; while $b_1 = \tilde{b}_1 + \ri a_1 /2$, thus leaving us with 4 real coefficients $a_1, a_2, a_5, \tilde{b}_1$.

In this particular case, the correlator is completely symmetric under point switches $1 \leftrightarrow 2$ and $2 \leftrightarrow 3$ (hence, it is also symmetric under $1 \leftrightarrow 3$). We must then impose further constraints due to these symmetries. The $1 \leftrightarrow 2$ symmetry demands that
\bea
H(X,P; \, u, \bar{u}, v, \bar{v}, w, \bar{w}) = H(-\bar{X}, P; \, v, \bar{v}, u, \bar{u}, w, \bar{w})~,
\eea
which translates to the following constraints
\bsubeq \label{12sym-111}
\bea
F(X; \, u, \bar{u}, v, \bar{v}, w, \bar{w}) &=& F(-X;  \,v, \bar{v}, u, \bar{u}, w, \bar{w})~, \\
G_{\a \ad}(X; \, u, \bar{u}, v, \bar{v}, w, \bar{w}) &=& G_{\a \ad}(-X;  \,v, \bar{v}, u, \bar{u}, w, \bar{w}) \non\\
&& +\,\ri \pa_{\a \ad} F(-X;\, v, \bar{v}, u, \bar{u}, w, \bar{w})~.~~~~~~~
\eea
\esubeq
Imposing \eqref{12sym-111} leads to $a_1=a_2=0$ and, hence, at this stage we end up with two real parameters: $a_5$ and $b_1$. The relations between coefficients are given by
\bea \label{coeffs-111}
&&a_1=a_2=a_3=a_4=0~, \non\\
&&b_2 = \frac{1}{4}(2 a_5 + b_1)~, ~~b_3= -3b_1~, ~~ b_4= b_5= -2 a_5 - \frac{3}{2}b_1~, ~~ b_6 = 2 a_5 + b_1~, \non\\
&& b_8 = \frac{1}{4}(2 a_5 + b_1)~, ~~ b_9= b_1~, ~~b_{10}= \frac{1}{4} (-2 a_5 + b_1)~, ~~b_{11}= -\frac{3}{2}b_1~, \\
&&b_{12}= -b_{13} =b_{14}= 2 \ri a_5~, ~~b_{15} = -\frac{3\ri}{2} a_5~.  \non
\eea
As for the $2 \leftrightarrow 3$ symmetry, it is not hard to show that the required constraints can be expressed in terms of \eqref{tildeF} and \eqref{tildeG} as follows:
\bsubeq 
\bea
\widetilde{F}(X; \, u, \bar{u}, v, \bar{v}, w, \bar{w}) &=& \widetilde{F}(-X;  \,v, \bar{v}, u, \bar{u}, w, \bar{w})~, \label{23sym-F} \\
\widetilde{G}_{\a \ad}(X; \, u, \bar{u}, v, \bar{v}, w, \bar{w}) &=& \widetilde{G}_{\a \ad}(-X;  \,v, \bar{v}, u, \bar{u}, w, \bar{w}) \non\\
&& +\,\ri \pa_{\a \ad} \widetilde{F}(-X;\, v, \bar{v}, u, \bar{u}, w, \bar{w})~.~~~~~~~ \label{23sym-G}
\eea
\esubeq
It can be verified that the $2 \leftrightarrow 3$ symmetry and conservation conditions at $z_3$ (described by eqs.~\eqref{conserv-3}--\eqref{conserv-3-end}) are automatically satisfied for the choice of coefficients \eqref{coeffs-111}. Our result here is thus in agreement with \cite{OsbornN1}. Indeed, our real parameters $a_5$ and $b_1$ are related to those in \cite{OsbornN1} by
\bea
a_5 = -A~, \qquad b_1 = 2(2C-A)~.
\eea

\subsection{Analysis for $r=2$}
We construct a generating function
\bea
&&H(X,P; \, u, \bar{u}, v, \bar{v}, w, \bar{w})
=\mathbf{U}^{\ad(2) \a(2) } \mathbf{V}^{\bd \b } \mathbf{W}^{\gd \g } H_{\a(2)\ad(2), \, \b \bd, \, \g \gd} (X, P) \non\\
&=& F (X;\, u, \bar{u}, v, \bar{v}, w, \bar{w}) -\hf P^{\ad \a} G_{\a \ad} (X;\, u, \bar{u}, v, \bar{v}, w, \bar{w})~.
\eea
The general expansion for the polynomial $F (X;\, u, \bar{u}, v, \bar{v}, w, \bar{w})$ is thus
\bea
F(X;\, u, \bar{u}, v, \bar{v}, w, \bar{w}) &=& \frac{1}{X^4}\cF(\hat{X}; u, \bar{u}, v, \bar{v}, w, \bar{w} )~.
\eea
Here the polynomial $\cF$ is homogeneous of degree 0 in $X$, degree 2 in $u, \bar{u}$ and degree 1 in $v, \bar{v}, w, \bar{w}$ (equivalently, we can say that it is of degree 2 in $U$ and degree 1 in $V$ and $W$). It is not difficult to construct all possible polynomial structures for $\cF$ in terms of our basis structures described in \eqref{basis-F}. Our expansion gives 6 independent structures:
\bea
\cF(\hat{X}; u, \bar{u}, v, \bar{v}, w, \bar{w} )&=& (W \hat{X}) \big[a_1 (U \hat{X})^2 (V \hat{X}) + a_2 (UV) (U \hat{X}) 
\big] \non\\
&+& {}(U W) \big[a_3 (U \hat{X}) (V \hat{X}) + a_4 (UV) \big] \non\\
&+& {} (V W) \big[a_5 (U \hat{X})^2\big] + a_6 \, (U \hat{X}) \cJ~.
\eea
We write $F(X;\, u, \bar{u}, v, \bar{v}, w, \bar{w}) = F_1(X;\, u, \bar{u}, v, \bar{v}, w, \bar{w})+ F_2(X;\, u, \bar{u}, v, \bar{v}, w, \bar{w})$, where
\bsubeq
\bea
F_1(X;\, u, \bar{u}, v, \bar{v}, w, \bar{w}) &= &\frac{1}{X^4} \Big[
a_1 (W \hat{X}) (U \hat{X})^2 (V \hat{X}) + a_2  (W \hat{X}) (UV) (U \hat{X}) \non\\
&&\qquad  +\,a_3 (U W) (U \hat{X}) (V \hat{X}) + a_4  (U W)(UV) \non\\
&&\qquad +\, a_5 (V W)(U \hat{X})^2
\Big]~,\\
F_2(X;\, u, \bar{u}, v, \bar{v}, w, \bar{w}) &=& \frac{a_6}{X^4} (U \hat{X}) \cJ~.
\eea
\esubeq
We shall now find the general expansion for the vector $G_{\a \ad}(X;\, u, \bar{u}, v, \bar{v}, w, \bar{w})$
\bea
G_{\a \ad}(X;\, u, \bar{u}, v, \bar{v}, w, \bar{w}) = \frac{1}{X^4} \cG_{\a \ad}(\hat{X};\, u, \bar{u}, v, \bar{v}, w, \bar{w})~,
\eea
and decompose $\cG$ further using the structures defined in \eqref{basis-G}:
\bea
\cG_{\a \ad}(\hat{X};\, u, \bar{u}, v, \bar{v}, w, \bar{w}) = \sum_{I=1}^{8} \cK_{I, \a \ad} \bm{\cG}_{I}(\hat{X};\, u, \bar{u}, v, \bar{v}, w, \bar{w})~.
\eea
In particular, we have that
\bsubeq
\bea
&&\text{$\cK_1$ structures:}\non\\
&&\qquad \quad \cK_{1, \a \ad} \bm{\cG}_{1}(\hat{X};\, u, \bar{u}, v, \bar{v}, w, \bar{w}) = u_{\a} \bar{u}_{\ad} \Big[ b_1 (U \hat{X}) (V \hat{X}) (W \hat{X}) + b_2 (U \hat{X}) (VW) \non\\
&&\hspace{6cm} +\, b_3 (V \hat{X}) (UW) + b_4 (W \hat{X}) (UV) +c_1 \cJ
\Big]~,\\
&&\text{$\cK_2$ structures:}\non\\
&&\qquad \quad \cK_{2, \a \ad} \bm{\cG}_{2}(\hat{X};\, u, \bar{u}, v, \bar{v}, w, \bar{w}) = \hat{X}_{\a \ad} \Big[ b_5 (U \hat{X})^2 (V \hat{X}) (W \hat{X}) + b_6 (U \hat{X})^2 (VW)  \non\\
&&\hspace{6cm} + b_7 (U \hat{X}) (V \hat{X}) (UW) + b_8 (U \hat{X}) (W \hat{X}) (UV) \non\\
&&\hspace{6cm} + b_9 (UV) (UW) + c_2  (U \hat{X}) \cJ \Big]~, \\
&&\text{$\cK_3$ structures:}\non\\
&&\qquad \quad \cK_{3, \a \ad} \bm{\cG}_{3}(\hat{X};\, u, \bar{u}, v, \bar{v}, w, \bar{w}) = v_{\a} \bar{v}_{\ad} \Big[ b_{10} (U \hat{X}) (UW)  + b_{11} (U\hat{X})^2 (W \hat{X}) \Big]~,~~~~~~~~\\
&&\text{$\cK_4$ structures:}\non\\
&&\qquad \quad \cK_{4, \a \ad} \bm{\cG}_{4}(\hat{X};\, u, \bar{u}, v, \bar{v}, w, \bar{w}) = w_{\a} \bar{w}_{\ad} \Big[ b_{12} (U \hat{X}) (UV)  + b_{13} (U \hat{X})^2 (V \hat{X}) \Big]~,~~~~~~~~\\
&&\text{$\cK_5$ structures:}\non\\
&&\qquad \quad \cK_{5, \a \ad} \bm{\cG}_{5}(\hat{X};\, u, \bar{u}, v, \bar{v}, w, \bar{w}) = \cZ_{1\, \a \ad} \Big[ b_{14} (U \hat{X}) (W \hat{X}) + b_{15} (UW) \Big]~,\\
&&\text{$\cK_6$ structures:}\non\\
&&\qquad \quad \cK_{6, \a \ad} \bm{\cG}_{6}(\hat{X};\, u, \bar{u}, v, \bar{v}, w, \bar{w}) = \cZ_{2\, \a \ad} \Big[ b_{16} (U \hat{X}) (V \hat{X}) + b_{17} (UV)\Big]~,\\
&&\text{$\cK_7$ structure:}\non\\
&&\qquad \quad \cK_{7, \a \ad} \bm{\cG}_{7}(\hat{X};\, u, \bar{u}, v, \bar{v}, w, \bar{w}) =\cZ_{3\, \a \ad} \big[ b_{18} (U \hat{X})^2 \big]~,\\ 
&&\text{$\cK_8$ structure:}\non\\
&&\qquad \quad \cK_{8, \a \ad} \bm{\cG}_{8}(\hat{X};\, u, \bar{u}, v, \bar{v}, w, \bar{w}) = \, \cZ_{4\, \a \ad} \big[b_{19}(U \hat{X}) \big] ~.
\eea
\esubeq
As in the previous subsection, we know that not all of the  above $\cK$-structures are linearly independent. In addition to the linear dependence relation \eqref{lin-dep-1}, we also have the following relation
\bea
u_{\a} \bar{u}_{\ad} \cJ + (U W) \cZ_{1\, \a \ad} - (UV)\cZ_{2\, \a \ad} - \cZ_{4\, \a \ad} (U \hat{X})= 0~. \label{lin-dep-2}
\eea  
Hence, we may choose to construct a linearly independent basis of polynomial structures for the vector $G_{\a \ad}(X;\, u, \bar{u}, v, \bar{v}, w, \bar{w})$ by removing the $c_{1}$ and $c_2$ structures.
This then allows us to write
\bsubeq
\bea
&&G_{\a \ad}(X;\, u, \bar{u}, v, \bar{v}, w, \bar{w}) = G_{1\,\a \ad}(X;\, u, \bar{u}, v, \bar{v}, w, \bar{w})+ G_{2\, \a \ad}(X;\, u, \bar{u}, v, \bar{v}, w, \bar{w})~,~~~~~~\\
&&G_{1\,\a \ad}(X;\, u, \bar{u}, v, \bar{v}, w, \bar{w}) = \frac{1}{X^5} \sum_{I=1}^{4} \cK_{I, \a \ad} \bm{\cG}_{I}(\hat{X};\, u, \bar{u}, v, \bar{v}, w, \bar{w})~,\\
&&G_{2\,\a \ad}(X;\, u, \bar{u}, v, \bar{v}, w, \bar{w}) = \frac{1}{X^5} \sum_{I=5}^{8} \cK_{I, \a \ad} \bm{\cG}_{I}(\hat{X};\, u, \bar{u}, v, \bar{v}, w, \bar{w})~.
\eea
\esubeq
We also note that all $a_i$ and $b_i$ are still assumed to be complex and hence, initially we have 25 complex parameters.

The next step is to impose differential constraints arising from the conservation on the first two points, eqs.~\eqref{conserv-1}--\eqref{conserv-2-end}. This can be computed relatively quickly using \textit{Mathematica}. These conditions fix the correlation function up to 4 independent complex coefficients, which we choose to be $a_1, a_5, a_6, b_1$. Specifically, the relations between the coefficients are given by:
\bea
&&a_2= - \frac{1}{6}a_1 -a_5~, \qquad  a_3 = \frac{1}{2}a_1+a_5~, \qquad a_4 = -\frac{1}{6}a_1~, \non\\
&&b_2 = -\frac{\ri}{4} \big( a_1-4a_5+4\ri a_6 +\ri b_1
\big)~, \hspace{0.5cm} \quad b_3 = \frac{1}{4} \big( 2\ri a_5+2 a_6 +b_1
\big)~,\non\\
&&b_4 = \frac{\ri}{12} \big(2 a_1 - 6 a_5+ 6 \ri a_6+3 \ri  b_1 \big)~, \non\\
&&b_5 = -2 \ri \big( a_1 - \ri b_1 \big)~, \hspace{1.5cm} 
b_6 =  \ri \big( a_1 - 3a_5+ \ri(3a_6+b_1) \big) ~, \non\\
&&b_7 = \frac{\ri}{2}\big( a_1 -6a_5+ 6\ri a_6 + 4 \ri b_1 \big)~,\non\\
&&b_8 = -\frac{\ri}{2}\big( a_1 -6a_5+ 6\ri a_6 + 2 \ri b_1 \big)~, \qquad \,\,\, b_{9} = \frac{1}{12} \big( \ri  a_1 + 3 b_1\big)~, \non\\
&&b_{10} = \frac{\ri}{4} \big( 2 a_5-2 \ri a_6 - \ri b_1 \big)~, \hspace{2.2cm} \, b_{11} = \frac{b_1}{2}~, \non\\
&&b_{12}= -\frac{\ri}{12} \big( 4 a_1 +6 a_5-6 \ri a_6+3 \ri b_1\big)~, \quad b_{13} = \frac{3\ri}{2}a_1-b_1~,\non\\
&&b_{14} = -\frac{a_1}{2}+ 3\ri(\ri a_5+a_6)~,\hspace{2cm} \quad b_{15} = \frac{1}{12} \big( a_1 -6 \ri( \ri a_5+ a_6) \big) ~,\non\\
&&b_{16} = \frac{3}{2} \big( a_1 -2\ri (\ri a_5+ a_6)
\big)~, \hspace{2.2cm}  b_{17} = - \frac{1}{4} \big(  a_1- 2\ri  (\ri a_5+ a_6) \big) ~,\non\\
&&b_{18} = -a_1+3 \ri (\ri a_5+a_6)~, \non\\
&&b_{19} = \frac{2}{3} a_1- 2 \ri (\ri a_5+a_6)~.
\eea
We must also impose the reality condition \eqref{real-all}, from which we obtain that $a_1, a_5, a_6$ are all real; while $b_1 = \tilde{b}_1 + \ri a_1$. Hence, we now have 4 real coefficients. 

Unlike the three-point function of the ordinary supercurrent, here we do not impose symmetry under $1 \leftrightarrow 2$. However, we do have symmetry under $2 \leftrightarrow 3$. It suffices to require that  the correlator be invariant under $2 \leftrightarrow 3$ and the conservation on $z_3$ will hold automatically. This computation again can be done in \textit{Mathematica} by first defining $\tilde{F}$ and $\tilde{G}_{\a \ad}$. Now, imposing \eqref{23sym-F} leads to $a_6 =0$, while the second condition \eqref{23sym-G} implies that $\tilde{b}_1=0$ and so $b_1 = \ri a_1$.  This then leaves us with just two independent real parameters: $a_1$ and $a_5$.

\subsection{Analysis for general $r > 2$}\label{subsect4.3}
We now complete the analysis for arbitrary $r > 2$.  
The general expansion for the polynomial $F (X;\, u, \bar{u}, v, \bar{v}, w, \bar{w})$ is thus
\bea
F(X;\, u, \bar{u}, v, \bar{v}, w, \bar{w}) &=& \frac{1}{X^{r+2}}\cF(\hat{X}; u, \bar{u}, v, \bar{v}, w, \bar{w} )~.
\eea
Our polynomial $\cF(\hat{X}; u, \bar{u}, v, \bar{v}, w, \bar{w} )$ is homogeneous of degree 0 in $X$, degree $r$ in $U$ and degree 1 in $V$ and $W$. This gives 6 independent structures, just like the $r=2$ case:
\bea
\cF(\hat{X}; u, \bar{u}, v, \bar{v}, w, \bar{w} )&=& (W \hat{X}) \big[a_1 (U \hat{X})^r (V \hat{X}) + a_2 (U \hat{X})^{r-1} (UV)  
\big] \non\\
&+& {}(U W) \big[a_3 (U \hat{X})^{r-1} (V \hat{X}) + a_4 (U \hat{X})^{r-2}(UV) \big] \non\\
&+& {} (V W) \big[a_5 (U \hat{X})^r\big] + a_6 \, (U \hat{X})^{r-1} \cJ~.
\eea
We then express 
\bsubeq
\bea
&&F(X;\, u, \bar{u}, v, \bar{v}, w, \bar{w}) = F_1(X;\, u, \bar{u}, v, \bar{v}, w, \bar{w})+ F_2(X;\, u, \bar{u}, v, \bar{v}, w, \bar{w})~,~~~~~~~\\
&&F_1(X;\, u, \bar{u}, v, \bar{v}, w, \bar{w}) = \frac{1}{X^{r+2}} \Big[
a_1 (W \hat{X}) (U \hat{X})^r (V \hat{X}) + a_2  (W \hat{X}) (UV) (U \hat{X})^{r-1} \non\\
&&\hspace{4.5cm}  +\,a_3 (U W) (U \hat{X})^{r-1} (V \hat{X}) + a_4  (U W) (U \hat{X})^{r-2} (UV) \non\\
&&\hspace{4.5cm} +\, a_5 (V W)(U \hat{X})^r
\Big]~,\\
&&F_2(X;\, u, \bar{u}, v, \bar{v}, w, \bar{w}) = \frac{a_6}{X^{r+2}} (U \hat{X})^{r-1} \cJ~.
\eea
\esubeq
As for the general expansion for $\cG_{\a \ad}(\hat{X};\, u, \bar{u}, v, \bar{v}, w, \bar{w})$, we write
\bea
\cG_{\a \ad}(\hat{X};\, u, \bar{u}, v, \bar{v}, w, \bar{w}) = \sum_{I=1}^{8} \cK_{I, \a \ad} \bm{\cG}_{I}(\hat{X};\, u, \bar{u}, v, \bar{v}, w, \bar{w})~.
\eea
In particular, we have that
\bsubeq
\bea
&&\text{$\cK_1$ structures:}\non\\
&&\quad \cK_{1, \a \ad} \bm{\cG}_{1}(\hat{X};\, u, \bar{u}, v, \bar{v}, w, \bar{w}) = u_{\a} \bar{u}_{\ad} \Big[ b_1 (U \hat{X})^{r-1} (V \hat{X}) (W \hat{X}) + b_2 (U \hat{X})^{r-1} (VW) \non\\
&&\hspace{5cm} +\, b_3 (U \hat{X})^{r-2} (V \hat{X}) (UW) + b_4 (U \hat{X})^{r-2}(W \hat{X}) (UV) \non\\
&&\hspace{5cm} +\, b_{20} (U \hat{X})^{r-3} (UV) (UW) + c_1 (U \hat{X})^{r-2} \cJ
\Big]~,\\
&&\text{$\cK_2$ structures:}\non\\
&&\quad \cK_{2, \a \ad} \bm{\cG}_{2}(\hat{X};\, u, \bar{u}, v, \bar{v}, w, \bar{w}) = \hat{X}_{\a \ad} \Big[ b_5 (U \hat{X})^r (V \hat{X}) (W \hat{X}) + b_6 (U \hat{X})^r (VW)  \non\\
&&\hspace{5cm} + b_7 (U \hat{X})^{r-1} (V \hat{X}) (UW) + b_8 (U \hat{X})^{r-1} (W \hat{X}) (UV) \non\\
&&\hspace{5cm} + b_{9} (U \hat{X})^{r-2}(UV) (UW) + c_2  (U \hat{X})^{r-1} \cJ \Big]~,\\
&&\text{$\cK_3$ structures:}\non\\
&&\quad \cK_{3, \a \ad} \bm{\cG}_{3}(\hat{X};\, u, \bar{u}, v, \bar{v}, w, \bar{w}) = v_{\a} \bar{v}_{\ad} \Big[ b_{10} (U \hat{X})^{r-1} (UW)  + b_{11} (U\hat{X})^r (W \hat{X}) \Big]~,~~~~~~~~\\
&&\text{$\cK_4$ structures:}\non\\
&&\quad \cK_{4, \a \ad} \bm{\cG}_{4}(\hat{X};\, u, \bar{u}, v, \bar{v}, w, \bar{w}) = w_{\a} \bar{w}_{\ad} \Big[ b_{12} (U \hat{X})^{r-1} (UV)  + b_{13} (U \hat{X})^r (V \hat{X}) \Big]~,~~~~~~~~\\
&&\text{$\cK_5$ structures:}\non\\
&&\quad \cK_{5, \a \ad} \bm{\cG}_{5}(\hat{X};\, u, \bar{u}, v, \bar{v}, w, \bar{w}) = \cZ_{1\, \a \ad} \Big[ b_{14} (U \hat{X})^{r-1} (W \hat{X}) + b_{15} (U \hat{X})^{r-2}(UW) \Big]~,~~~~~~~~\\
&&\text{$\cK_6$ structures:}\non\\
&&\quad \cK_{6, \a \ad} \bm{\cG}_{6}(\hat{X};\, u, \bar{u}, v, \bar{v}, w, \bar{w}) = \cZ_{2\, \a \ad} \Big[ b_{16} (U \hat{X})^{r-1} (V \hat{X}) + b_{17} (U \hat{X})^{r-2} (UV)\Big]~,~~~~~~~~\\
&&\text{$\cK_7$ structure:}\non\\
&&\quad \cK_{7, \a \ad} \bm{\cG}_{7}(\hat{X};\, u, \bar{u}, v, \bar{v}, w, \bar{w}) =\cZ_{3\, \a \ad} \big[ b_{18} (U \hat{X})^r \big]~,\\ 
&&\text{$\cK_8$ structure:}\non\\
&&\quad \cK_{8, \a \ad} \bm{\cG}_{8}(\hat{X};\, u, \bar{u}, v, \bar{v}, w, \bar{w}) = \, \cZ_{4\, \a \ad} \big[b_{19}(U \hat{X})^{r-1} \big] ~.
\eea
\esubeq
Here we note the additional $\cK_1$ structure containing the $(U\hat{X})^{r-3}$ term, which exists for $r \geq 3$.  Due to the linear dependence relations \eqref{lin-dep-1} and \eqref{lin-dep-2}, we may choose to construct a linearly independent basis of polynomial structures for $G_{\a \ad}(X;\, u, \bar{u}, v, \bar{v}, w, \bar{w})$ by removing the $c_1$ and $c_2$ structures. As a result, we can express
\bsubeq
\bea
&&G_{\a \ad}(X;\, u, \bar{u}, v, \bar{v}, w, \bar{w}) = G_{1\,\a \ad}(X;\, u, \bar{u}, v, \bar{v}, w, \bar{w})+ G_{2\, \a \ad}(X;\, u, \bar{u}, v, \bar{v}, w, \bar{w})~,~~~~~~~\\
&&G_{1\,\a \ad}(X;\, u, \bar{u}, v, \bar{v}, w, \bar{w}) = \frac{1}{X^{r+3}} \sum_{I=1}^{4} \cK_{I, \a \ad} \bm{\cG}_{I}(\hat{X};\, u, \bar{u}, v, \bar{v}, w, \bar{w})~,\\
&&G_{2\,\a \ad}(X;\, u, \bar{u}, v, \bar{v}, w, \bar{w}) = \frac{1}{X^{r+3}} \sum_{I=5}^{8} \cK_{I, \a \ad} \bm{\cG}_{I}(\hat{X};\, u, \bar{u}, v, \bar{v}, w, \bar{w})~.
\eea
\esubeq
Since $a_i$ and $b_i$ are initially assumed to be complex, we have 26 complex parameters.

The six constraints originating from the conservation on the first two points, eqs.~\eqref{conserv-1}--\eqref{conserv-2-end} fix the correlation function up to 4 independent complex coefficients, which we choose to be $a_1, a_5, a_6, b_1$. For completeness, here we give the relations between them
\bea
&&a_2= - \frac{r-1}{2(r+1)}a_1 -a_5~, \qquad  a_3 = \frac{r-1}{2}a_1+a_5~, \qquad a_4 = -\frac{r(r-1)}{4(r+1)}a_1~, \non\\
&&b_2 = -\frac{\ri}{8} \Big[ r(a_1-4a_5+4\ri a_6)+2\ri b_1
\Big]~, \hspace{0.5cm} \quad b_3 = \frac{r-1}{8} \Big[ \ri(r-2) a_1+ 4 \big(\ri a_5 + a_6 \big) +2b_1
\Big]~,\non\\
&&b_4 = -\frac{r-1}{8(r+1)} \Big[-\ri(r+2) a_1 +4(r+1) (\ri a_5+a_6)+6b_1 \Big]~, \non\\
&&b_5 = -\frac{r+2}{2r} \Big[ \ri r a_1 + 2b_1 \Big]~, \hspace{1.5cm} b_6 = \frac{1}{4r}(r+2) \big( \ri r a_1- 2b_1 \big)-(r+1) \big(\ri a_5+a_6 \big) ~, \non\\
&&b_7 = -\frac{\ri}{4}(r^2-2r-2) a_1 -(r+1) \big(\ri a_5+a_6 \big)-\frac{r+2}{2} b_1 ~,\non\\
&&b_8 = -\frac{1}{2} \Big[\ri a_1 -2 (r+1) \big( \ri a_5+a_6 \big)-2 b_1 \Big]~, \quad ~b_{9} = \frac{(r-1)}{8(r+1)} \Big[ \ri r(r-1) a_1 + 2(r+1) b_1\Big]~, \non\\
&&b_{10} = \frac{1}{8} \Big[ \ri(r-2) a_1+ 4(\ri a_5+a_6)  +2b_1\Big]~, \qquad b_{11} = \frac{b_1}{r}~, \non\\
&&b_{12}= -\frac{1}{8} \Big[ \frac{\ri}{r+1}(r^2+3r-2)a_1 + 4(\ri a_5+a_6) -2 b_1 \Big]~, \quad b_{13} = \frac{\ri}{4}(r+4)a_1-\frac{r+2}{2r} b_1~,\non\\
&&b_{14} = -\hf \Big[ (r-1)a_1+2(r+1)(a_5- \ri a_6)
\Big]~,\non\\
&&b_{15} = \frac{r-1}{4(r+1)} \Big[ (r-1) a_1 -2 \ri(r+1) ( \ri a_5+ a_6) \Big] ~,\non\\
&&b_{16} = \frac{1}{2}(2r-1)a_1 -\ri (r+1) \big( \ri a_5 +a_6 \big)~, \non\\
&& b_{17} = - \frac{r-1}{4(r+1)} \Big[ (2r-1) a_1- 2\ri (r+1) (\ri a_5+ a_6) \Big] ~,\non\\
&&b_{18} = -\frac{r}{2}a_1+ \ri (r+1) \big(\ri a_5+ a_6\big)~, \non\\
&&b_{19} = \frac{1}{4} \Big[\frac{r^2+3r-2}{r+1}a_1 - \ri (2r+4)\big(\ri a_5 + a_6 \big)
\Big]~, \non\\
&&b_{20} = -\frac{(r-1)(r-2)}{16(r+1)} \big( \ri r a_1+2 b_1 \big)~.
\eea
We must also impose the reality condition \eqref{real-all}, from which we obtain that $a_1, a_5, a_6$ are all real; while 
\bea
b_1 = \tilde{b}_1 + \frac{\ri r }{2} a_1~.
\eea
Hence, at this stage we have 4 independent real coefficients. 

The constraints imposed by the symmetry under permutation of superspace points $z_2 \leftrightarrow z_3$ imply that the final form of the solution depends on 
whether $r$ is even or odd. More precisely, if $r$ is \textbf{even}, conditions \eqref{23sym-F} and \eqref{23sym-G} require that
\bea
a_6=0~, \qquad  ~\tilde{b}_1=0 \Longrightarrow~b_1 = \frac{\ri r}{2} a_1~,
\eea
so, the independent parameters are $a_1$ and $a_5$. This is consistent with what we found in the $r=2$ case. On the other hand, if $r$ is \textbf{odd}, the $2 \leftrightarrow 3$ symmetry implies that
\bea
a_1 = a_5 =0~,
\eea
so, the independent parameters are $a_6$ and $b_1$. 
Thus, in any case the correlator is still determined up to two real parameters, though its explicit structure is different.

\section{Correlators involving flavour current multiplet}
\label{section5}
We now turn to computing mixed correlators containing the higher-spin supercurrent with the flavour current multiplet. In $\cN=1$ superconformal field theory, the latter is described by a primary real scalar superfield $L$ of weights $(q, \bar{q})= (1,1)$ and dimension 2. It satisfies the conservation law
\bea
D^2 L = \bar{D}^2 L=0~.
\eea
\subsection{Correlator $\la L(z_1) L(z_2) J_{\a(r) \ad(r)} (z_3)\ra$}
By virtue of \eqref{3ptgen}, the ansatz for this three-point function takes the form
\bea 
\la L(z_1) L(z_2) J_{\a(r) \ad(r)} (z_3)\ra 
=\frac{1}{(x_{\bar{1} 3} x_{\bar{3} 1} x_{\bar{2} 3} x_{\bar{3} 2})^2}  H_{\a(r) \ad(r)} (\fn3)~.~~~~~~~~
\eea 
The tensor $H_{\a(r) \ad(r)}$ is subject to the following constraints:
\begin{enumerate}
	\item[\textbf{(i)}] \textbf{Homogeneity:} It has the scaling property
	\bea
	H_{\a(r) \ad(r)} (\L \bar{\L} X, \L \Q, \bar{\L} \bar{\Q}) 
	= (\L \bar{\L})^{r-2} H_{\a(r) \ad(r)} (\fxq)~, 	
	\eea
and, hence, its dimension is $(r-2)$. 
	\item[\textbf{(ii)}] \textbf{Conservation:} Conservation laws of $L$ on the first two points require that
\bsubeq
\bea
D^{2}_{(1)} \la L(z_1) L(z_2) J_{\a(r) \ad(r)} (z_3)\ra &=& 0~, \\
\bar D^{2}_{(1)} \la L(z_1) L(z_2) J_{\a(r) \ad(r)} (z_3)\ra &=& 0~,\\
D^{2}_{(2)}  \la L(z_1) L(z_2) J_{\a(r) \ad(r)} (z_3)\ra &=& 0~, \\
\bar D^{2}_{(2)} \la L(z_1) L(z_2) J_{\a(r) \ad(r)} (z_3)\ra &=& 0~. 
\eea
\esubeq
With the use of \eqref{Cderivs}, these requirements are equivalent to imposing these differential constraints on $H$:
\bsubeq 
\bea
&&\bar{\cD}^{2} H_{\a(r) \ad(r)} = 0~,\label{LLJ-ceq1}\\
&&{\cD}^{2} H_{\a(r) \ad(r)} = 0~, \label{LLJ-ceq2} \\
&&\bar{\cQ}^{2} H_{\a(r) \ad(r)} = 0~, \label{LLJ-ceq3}\\
&&\cQ^{2}H_{\a(r) \ad(r)} = 0~.\label{LLJ-ceq4}
\eea
\esubeq
Eqs.~\eqref{LLJ-ceq1} and \eqref{LLJ-ceq4} tell us that again, the general solution for $H$ takes the form
\bea
H_{\a(r) \ad(r)} (X, \Q, \bar{\Q}) = H_{\a(r) \ad(r)} (X, P)= F_{\a(r) \ad(r)} (X) -\hf P^{\gd \g} G_{\g \gd, \,\a(r) \ad(r)} (X)~.~~~~~
\eea
This means that we can adopt a similar procedure as in section \ref{section3} to solve for the differential constraints in terms of polynomials $F(X;\, u, \bar{u})$ and $G_{\g \gd} (X;\, u,\bar{u})$:
\bsubeq
\bea
F(X;\, u, \bar{u}) &=& u^{\a_1} \dots u^{\a_r} \bar{u}^{\ad_1} \dots \bar{u}^{\ad_r} F_{\a(r) \ad(r)} (X)~, \\
G_{\g \gd}(X;\, u, \bar{u}) &=& u^{\a_1} \dots u^{\a_r} \bar{u}^{\ad_1} \dots \bar{u}^{\ad_r} G_{\g \gd,\,\a(r) \ad(r)} (X)~.
\eea
\esubeq
It is straightforward to check that eqs.~\eqref{LLJ-ceq2} and \eqref{LLJ-ceq3} both imply
\bea
\Box F(X;\, u, \bar{u})-\ri \pa^{\gd \g} G_{\g \gd}(X;\, u, \bar{u}) = 0~, \label{LLJ-conserv-FG}
\eea
which, in fact, is the consistency condition \eqref{consistency}.
We must also take into account the conservation of $J_{\a(r) \ad(r)}$,
\bea
D^{\b}_{(3)} \la L(z_1) L(z_2) J_{\b \a(r-1) \ad(r)} (z_3)\ra &=& 0~, \\
\bar D^{\bd}_{(3)} \la L(z_1) L(z_2) J_{\a(r) \bd \ad(r-1)} (z_3)\ra &=& 0~.
\eea
As usual, we can rearrange our correlator as
\bea
&&\la L(z_1) L(z_2) J_{\a(r) \ad(r)} (z_3)\ra = \la  L(z_2) J_{\a(r) \ad(r)} (z_3) L(z_1) \ra \non\\
&=& \frac{1}{(x_{\bar{1} 2} x_{\bar{2} 1} x_{\bar{1} 3} x_{\bar{3} 1})^2} \cI_{\a(k) \bd(k)} (x_{3 \bar{1}})  \cI_{\b(k) \ad(k)} (x_{1 \bar{3}}) \widetilde{H}^{\bd(k) \b(k)} (X_1, P_1)~,
\eea 
which, after some manipulations, leads to
\bea \label{LLJ-tildeH-HI}
&&\widetilde{H}_{\b(r) \bd(r)} (X_1, P_1) =\frac{\big(\l \bar{\l}\big)^{-(r-2)}}{\big( X_1 \bar{X}_1 \big)^r} \cI_{\a(r) \bd(r)} (x_{3 \bar{1}}) \cI_{\b(r) \ad(r)} (x_{1 \bar{3}})\, H^{\ad(r) \a(r)} (X_3, P_3) \non\\
&&= \frac{(-1)^r}{\big( X_1 \bar{X}_1 \big)^r} \Big[ F_{\b(r) \bd(r)} (X_1{}^{I}) -\hf P_1{}^{I, \, \ad \a} G_{\a \ad,\, \b(r) \bd(r)} (X_1{}^{I})\Big] \\
&&\equiv \frac{(-1)^r}{\big( X_1 \bar{X}_1 \big)^r} \,H_{\b(r) \bd(r)} (X_1{}^I, P_1{}^I)~.\non
\eea
As we shall see soon, the simple expression in \eqref{LLJ-tildeH-HI} is due to the fact that none of the structures in $H_{\a(r) \ad(r)} (X_3,P_3)$ transforms as a pseudotensor under the actions of $\cI_{\a(r) \bd(r)} (x_{3 \bar{1}}) \cI_{\b(r) \ad(r)} (x_{1 \bar{3}})$. 
Next, we can easily expand $\bar{X} = X+ \ri P$ and then start contracting with auxiliary spinors. This results in
\bsubeq \label{LLJ-tildeFG}
\bea
\widetilde{H}(X,P;\, u, \bar{u}) = \widetilde{F}(X;\, u, \bar{u}) -\hf P^{\ad \a} \widetilde{G}_{\a \ad}(X;\, u, \bar{u})~,
\eea
where
\bea
\widetilde{F}(X;\, u, \bar{u}) &=& \frac{1}{X^{2r}} F(X;\, u, \bar{u})~, \\
\widetilde{G}_{\a \ad}(X;\, u, \bar{u}) &=& \frac{\ri}{X^{2r}} \pa_{\a \ad} F(X;\, u, \bar{u}) -2\ri (r-1) \frac{X_{\a \ad}}{X^{2r+2}}F(X;\, u, \bar{u}) \non\\
&&-\frac{X_{\a \bd} X_{\b \ad}}{X^{2r+2}}  G^{\bd \b}(X;\, u, \bar{u})~.
\eea
\esubeq
Conservation of $J_{\a(r) \ad(r)}$ at $z_3$ then amounts to the following constraints:
\bsubeq \label{LLJ-conserv-z3}
\bea
&&\ve^{\a \b} \frac{\pa}{\pa u^{\b}} \widetilde{G}_{\a \ad} (X;\, u, \bar{u})= 0~,\\
&&\pa^{\bd \a} \frac{\pa}{\pa \bar{u}^{\bd}} \widetilde{G}_{\a \ad} (X;\, u, \bar{u}) = 0~,\\
&&\ve^{\ad \bd} \frac{\pa}{\pa \bar{u}^{\bd}} \Big[ \widetilde{G}_{\a \ad} (X;\, u, \bar{u}) - \ri \pa_{\a \ad} \widetilde{F} (X;\, u, \bar{u}) \Big] = 0~.
\eea
\esubeq

\item[\textbf{(iii)}] \textbf{Reality:} Since both $L$ and $J_{\a(r) \ad(r)}$ are real superfields, the reality condition on the correlator leads to
\bea
H(X,P;\, u, \bar{u};\, a_i, b_i) = \bar{H} (X,P;\, u, \bar{u};\, \bar{a}_i, \bar{b}_i)~,
\eea
which are equivalent to 
\bsubeq
\bea
F(X;\, u, \bar{u};\, a_i) &=& F(X;\, u, \bar{u};\, \bar{a}_i) \Longrightarrow a_i ~\rm{real}~, \\
G_{\g \gd} (X;\, u, \bar{u};\, b_i) &=& G_{\g \gd}(X;\, u, \bar{u};\, \bar{b}_i)+ \ri \pa_{\g \gd} F(X;\, u, \bar{u};\, \bar{a}_i)~. \label{LLJ-reality-G}
\eea
\esubeq

\item[\textbf{(iv)}] \textbf{$1 \leftrightarrow 2$ symmetry:} The correlator is symmetric under $z_1 \leftrightarrow z_2$ which imposes the following condition on $H$
\bea
H(X,P;\, u, \bar{u}) = H (-\bar{X},P;\, u, \bar{u})~,
\eea
or, equivalently,
\bsubeq \label{LLJ-12-all}
\bea
F(X;\, u, \bar{u}) &=& F(-X;\, u, \bar{u}) ~,\\
G_{\g \gd} (X;\, u, \bar{u}) &=& G_{\g \gd}(-X;\, u, \bar{u})+ \ri \pa_{\g \gd} F(-X;\, u, \bar{u})~. 
\eea
\esubeq
\end{enumerate}

We are now ready to construct the general structures for $F(X;\, u, \bar{u})$ and $G_{\g \gd}(X;\, u, \bar{u})$. In terms of our invariants, there is only one possible structure for $F$:
\bea
F(X;\, u, \bar{u}) &=& \frac{a_1}{X^{2-r}} (U \hat{X} )^r \non\\
&=&  \Big(-\hf \Big)^{r} \frac{a_1}{X^{2-r}} \hat{X}_{(u, \bar{u})}^r~.
\eea
There are two independent structures for $G_{\g \gd} (X; \, U)$:
\bea
G_{\g \gd}(X;\, u, \bar{u}) &=& \frac{1}{X^{3-r}}  \Big[ b_1 \,\hat{X}_{\g \gd}(U \hat{X} )^r + b_2 \,u_{\g}\bar{u}_{\gd} (U \hat{X})^{r-1} \Big] \non\\
&=&  \Big(-\hf \Big)^r \frac{1}{X^{3-r}}  \Big[ b_1 \hat{X}_{\g \gd}\hat{X}_{(u,\bar{u})}^r -2 b_2 \,u_{\g}\bar{u}_{\gd} \hat{X}_{(u, \bar{u})}^{r-1} \Big]~.
\eea
The conservation condition \eqref{LLJ-conserv-FG} gives
\bea
b_2 = \frac{r}{2} (b_1 + 2 \ri a_1)~, \label{b2-relation}
\eea
where, at this stage the coefficients $a_1$ and $b_1$ are still assumed to be complex. 
Next, upon substituting the explicit expressions of $F(X;\, u, \bar{u})$ and $G_{\g \gd}(X;\, u, \bar{u})$ into \eqref{LLJ-tildeFG}, one can verify that conservation conditions at $z_3$, eqs.\eqref{LLJ-conserv-z3}, are identically satisfied if we impose \eqref{b2-relation}. Reality conditions imply that $a_1$ is real, while \eqref{LLJ-reality-G} gives
\bea
b_1 - \bar{b}_1 = -2\ri a_1~.
\eea
Finally, the $1 \leftrightarrow 2$ symmetry \eqref{LLJ-12-all} imposes the following relations
\bea
a_1 = (-1)^r a_1 = \frac{\ri}{2} b_1 \big( (-1)^r +1 \big)~.
\eea
Thus, for odd values of $r = 2k+1$, we have that $a_1=0~, \bar{b}_1 = b_1$, while for even values of $r = 2k$, the coefficients are related by $b_1 = -\ri a_1$. 

Our final result is that the correlator $\la L(z_1) L(z_2) J_{\a(r) \ad(r)} (z_3)\ra$ is fixed up to a single real parameter, though its explicit form depends on the values of $r$:
\begin{itemize}
\item \textbf{$r$ odd:}
\bsubeq
\bea
H(X, P;\, u, \bar{u}) = -\hf P^{\ad \a} G_{\a \ad} (X;\, u, \bar{u})~,
\eea
with 
\bea
G_{\a \ad} (X; u, \bar{u}) = \frac{b_1}{X^{3-r}}  \Big[\hat{X}_{\a \ad}\hat{X}_{(u,\bar{u})}^r - r \,u_{\a}\bar{u}_{\ad} \,\hat{X}_{(u, \bar{u})}^{r-1} \Big]~.
\eea
\esubeq
\item \textbf{$r$ even:}
\bsubeq
\bea
H(X, P;\, u, \bar{u}) &=& F(X; \, u, \bar{u}) -\hf P^{\ad \a} G_{\a \ad} (X;\, u, \bar{u})~,
\eea
with 
\bea
F(X; \, u, \bar{u}) &=& \frac{a_1}{X^{2-r}} \hat{X}_{(u, \bar{u})}^r~,\\
G_{\a \ad} (X; u, \bar{u}) &=& -\frac{\ri a_1}{X^{3-r}}  \Big[\hat{X}_{\a \ad}\hat{X}_{(u,\bar{u})}^r + r \,u_{\a}\bar{u}_{\ad} \,\hat{X}_{(u, \bar{u})}^{r-1} \Big]~.
\eea
\esubeq
\end{itemize}

\subsection{Correlator $\la J_{\a(r_1) \ad(r_1)}(z_1) J_{\b(r_2) \bd(r_2)}(z_2) L(z_3)\ra$}
Using the general prescription \eqref{3ptgen}, the ansatz for this correlator is given by
\bea 
\la J_{\a(r_1) \ad(r_1)}(z_1)  J_{\b(r_2) \bd(r_2)}(z_2) L(z_3) \ra 
&=&
\frac{1}{(x_{\bar{1} 3} x_{\bar{3} 1})^{r_1+2} (x_{\bar{2} 3} x_{\bar{3} 2})^{r_2+2}}  \cI_{\a(r_1) \gd(r_1)} (x_{1 \bar{3}})\, \cI_{\a(r_1) \gd(r_1)} (x_{3 \bar{1}}) \non\\
&&\quad \times~ \cI_{\b(r_2) \dot{\d}(r_2)} (x_{2 \bar{3}})\, \cI_{\d(r_2) \bd(r_2)} (x_{3 \bar{2}}) \non\\
&&\quad \times~ H^{\gd(r_1) \g(r_1),\,\, \dot{\d}(r_2)\d(r_2)} (\fn3)~.~~~~~~~~
\eea 
The tensor $H_{\a(r_1) \ad(r_1),\,\, \b(r_2) \bd(r_2)}$ is subject to the following constraints:
\begin{enumerate}
	\item[\textbf{(i)}] \textbf{Homogeneity:} It is characterised by the scaling property
	\bea
	&& H_{\a(r_1) \ad(r_1),\,\, \b(r_2) \bd(r_2)} (\L \bar{\L} X, \L \Q, \bar{\L} \bar{\Q}) \non\\
	&=& (\L \bar{\L})^{-(r_1+r_2+2)} H_{\a(r_1) \ad(r_1),\,\, \b(r_2) \bd(r_2)} (\fxq)~, 	
	\eea
and, hence, its dimension is $-(r_1+r_2+2)$. 
	\item[\textbf{(ii)}] \textbf{Conservation:} Conservation conditions on the first two points require
\bsubeq 
\bea
&&\bar{\cD}^{\gd} H_{\a(r_1) \gd \ad(r_1-1),\,\, \b(r_2) \bd(r_2)} = 0~,\label{JJL-ceq1}\\
&&{\cD}^{\g} H_{\g \a(r_1-1) \ad(r_1),\,\, \b(r_2) \bd(r_2)} = 0~, \label{JJL-ceq2} \\
&&\cQ^{\g}H_{\a(r_1) \ad(r_1),\,\, \g \b(r_2-1) \bd(r_2)}=0~,\label{JJL-ceq3}\\
&&\bar{\cQ}^{\gd} H_{\a(r_1) \ad(r_1),\,\, \b(r_2) \gd \bd(r_2-1)}
= 0~.\label{JJL-ceq4}
\eea
\esubeq
Furthermore, eqs.~\eqref{JJL-ceq1} and \eqref{JJL-ceq3} also suggest that the general solution for $H$ is
\bea
H(X,P; \, u, \bar{u}, v, \bar{v})
&=& \mathbf{U}^{\ad(r_1) \a(r_1) } \mathbf{V}^{\bd(r_2) \b(r_2) }  H_{\a(r_1) \ad(r_1),\,\, \b(r_2) \bd(r_2)} (X, P) \non\\
&=& F (X;\, u, \bar{u}, v, \bar{v}) -\hf P^{\ad \a} G_{\a \ad} (X;\, u, \bar{u}, v, \bar{v})~,~~~~~~
\eea
where, in the above we have contracted all indices with the commuting spinors. Here the polynomials $F (X;\, u, \bar{u}, v, \bar{v})$ and $G_{\a \ad} (X;\, u, \bar{u}, v, \bar{v})$ satisfy the differential constraints given by eqs.~\eqref{conserv-1}--\eqref{conserv-2-end}. Conservation conditions on $z_3$ will be worked out at the end, once we know the resulting tensorial structures for $F$ and $G_{\a \ad}$ after imposing conservations on the first two points and the reality constraint. As can be seen from \eqref{tildeF}--\eqref{tildeG} (and eqs.~\eqref{LLJ-tildeFG} for comparison), the expression for $\widetilde{H}$ will be greatly simplified if no $\e_{abcd}$ terms are present in the solution. 

\item[\textbf{(iii)}] \textbf{Reality:} The reality condition on the correlator leads to
\bea
H(X,P;\, u, \bar{u}, v, \bar{v};\, a_i, b_i) = \bar{H} (X,P;\, u, \bar{u}, v, \bar{v};\, \bar{a}_i, \bar{b}_i)~,
\eea
which are equivalent to 
\bsubeq
\bea
F(X;\, u, \bar{u}, v, \bar{v};\, a_i) &=& F(X;\, u, \bar{u}, v, \bar{v};\, \bar{a}_i) \Longrightarrow a_i ~\rm{real}~, \\
G_{\a \ad} (X;\, u, \bar{u}, v, \bar{v};\, b_i) &=& G_{\a \ad}(X;\, u, \bar{u}, v, \bar{v};\, \bar{b}_i)+ \ri \pa_{\a \ad} F(X;\, u, \bar{u}, v, \bar{v};\, \bar{a}_i)~.~~~~~~ \label{JJL-reality-G}
\eea
\esubeq
\end{enumerate}

\subsubsection{Analysis for $1\leq r_1<r_2$}
Without loss of generality, let us assume that $r_1 < r_2$ and so $ {\rm min} (r_1, r_2) = r_1$. In this case we will not impose invariance under the exchange of superspace points $z_1 \leftrightarrow z_2$.
We proceed with constructing the tensorial structure for $F(X;\, u, \bar{u}, v, \bar{v})$. 
\bea
F(X;\, u, \bar{u}, v, \bar{v}) = \frac{1}{X^{r_1+r_2+2}} \sum_{k_i} a(k_i) (U \hat{X})^{k_1} (V \hat{X})^{k_2} (UV)^{k_3}~,
\eea
where the degrees of homogeneity in $U$ and $V$ imply that 
\bea
k_1+ k_3 = r_1~, \qquad 
k_2+ k_3 =r_2~.
\eea
In addition, it holds that $k_i \geq 0$. These requirements allow us to write
\bea
F(X;\, u, \bar{u}, v, \bar{v}) = \frac{1}{X^{r_1+r_2+2}} \sum_{k=0}^{r_1} a_k \,(U \hat{X})^{r_1-k} (V \hat{X})^{r_2-k} (UV)^{k}~.
\label{JJL-F}
\eea
As for the structure of $G_{\a \ad}(X;\, u, \bar{u}, v, \bar{v})$, we can write
\bsubeq \label{JJL-G}
\bea
G_{\a \ad}(X;\, u, \bar{u}, v, \bar{v}) = \frac{1}{X^{r_1+r_2+3}} \cG_{\a \ad}(\hat{X};\, u, \bar{u}, v, \bar{v})~,
\eea
with
\bea
\cG_{\a \ad}(\hat{X};\, u, \bar{u}, v, \bar{v}) &=& 
\hat{X}_{\a \ad}\sum_{k=0}^{r_1} B_k \,(U \hat{X})^{r_1-k} (V \hat{X})^{r_2-k} (UV)^{k}  \non\\
&+& u_{\a} \bar{u}_{\ad} \sum_{k=0}^{r_1-1} C_k \,(U \hat{X})^{r_1-(k+1)} (V \hat{X})^{r_2-k} (UV)^{k}  \\
&+& v_{\a} \bar{v}_{\ad} \sum_{k=0}^{r_1} D_k \,(U \hat{X})^{r_1-k} (V \hat{X})^{r_2-(k+1)} (UV)^{k}  \non\\
&+& {\cZ}_{1 \,\a \ad}\sum_{k=0}^{r_1-1} E_k \,(U \hat{X})^{r_1-(k+1)} (V \hat{X})^{r_2-(k+1)} (UV)^{k}~,\non
\eea
\esubeq
where we recalled that $\cZ_{1 \, \a \ad} = (\s^a)_{\a \ad} \cZ_{1\, a} = (\s^a)_{\a \ad} \, \e_{abcd} \hat{X}^b U^c V^d$. Thus, initially we have $(5r_1+3)$ independent, complex parameters. To further constrain these parameters, we impose conservation conditions on the first two points, eqs.~\eqref{conserv-1}--\eqref{conserv-2-end}. Here one can verify that $E_k =0$, for $k=0,1, \dots r_1-1$, which will certainly simplify our calculation for $\widetilde{H}$ later as there is no need to split $G_{\a \ad}$ into two sectors (tensor and pseudotensor ones). Taking the reality condition into account, we find that, for $k=0, 1, \dots, r_1$:
\bsubeq \label{JJL-parameters}
\bea
B_k &=& b_k -\ri \,(r_1+r_2+1-k) \,a_k~,\\
C_k &=& -\,\frac{k-r_1}{2(r_1+r_2+1-k)} \,b_k + \frac{\ri}{2} (r_1-k) \,a_k~,\\
D_k &=& -\,\frac{r_2-k}{2(r_1+r_2+1-k)}\, b_k + \frac{\ri}{2} (r_2-k)\, a_k~,
\eea
where the parameters $a_k$ and $b_k$ are real. Furthermore, for $k=1, 2, \dots r_1$, they obey following recursive relations:
\bea
a_k &=& -\frac{(r_1+1-k) (r_2+1-k)}{2k(r_1+r_2+1-k)} a_{k-1}~, \\
b_k &=& -\frac{(r_1+1-k) (r_2+1-k)}{2k(r_1+r_2+2-k)} b_{k-1}~.
\eea
\esubeq
The above relations imply that we are now left with just two real parameters: $a_0$ and $b_0$. 

Let us analyse the constraints resulting from the conservation law of $L$ on $z_3$. Rewriting our correlator as
\bea
&&\la  J_{\b(r_2) \bd(r_2)}(z_2) L(z_3) J_{\a(r_1) \ad(r_1)}(z_1) \ra \non\\
&=& \frac{1}{(x_{\bar{1} 3} x_{\bar{3} 1})^2 (x_{\bar{1} 2} x_{\bar{2} 1})^{r_2+2}}~ \cI_{\b(r_2) \gd(r_2)} (x_{2 \bar{1}}) \, \cI_{\g(r_2) \bd(r_2)} (x_{1 \bar{2}}) \,\widetilde{H}^{\gd(r_2) \g(r_2)}{}_{
\a(r_1) \ad(r_1)}(X_1, P_1)~,~~~~~~
\eea
and making use several identities analogous to \eqref{x-F} and \eqref{x-G}, we obtain
\bea
&&\widetilde{H}_{\b(r_2) \bd(r_2),\, \a(r_1) \ad(r_1)} (X, P)\non\\
&=& (-X \bar{X})^{r_1} \, \cI_{\b(r_2)}{}^{\gd(r_2)} (X) \, \cI_{\bd(r_2)}{}^{\g(r_2)}(\bar{X})\, H_{\a(r_1) \ad(r_1),\, \g(r_2) \gd(r_2)} (X^I, P^I)~.
\eea
Defining the polynomial $\widetilde{H} (X, P;\, u, \bar{u}, v, \bar{v})$  by
\bea
\widetilde{H} (X, P;\, u, \bar{u}, v, \bar{v}) =  \mathbf{U}^{\bd(r_2) \b(r_2) } \mathbf{V}^{\ad(r_1) \a(r_1) } \widetilde{H}_{\b(r_2) \bd(r_2),\, \a(r_1) \ad(r_1)} (X, P)~,
\eea
we obtain the relation
\bea \label{JJL-tildeH-I}
\widetilde{H} (X, P;\, u, \bar{u}, v, \bar{v}) = \frac{(-1)^{r_1}}{r_2! r_2!} (X \bar{X})^{r_1-r_2} 
\big(\bar{u} \bar{X} \pa_{(s)} \big)^{r_2}\big( {u} {X} \pa_{(\bar{s})} \big)^{r_2}\, H(X^I, P^I;\, v, \bar{v}, s, \bar{s})~. ~~~~~
\eea
Here $H(X^I, P^I;\, v, \bar{v}, s, \bar{s})$ is defined by
\bea
&&H(X^I, P^I;\, v, \bar{v}, s, \bar{s}) = F(X^I;\, v, \bar{v}, s, \bar{s}) -\hf P^{I,\, \ad \a} G_{\a \ad} (X^I;\, v, \bar{v}, s, \bar{s}) \non\\
&=& F(X^I;\, v, \bar{v}, s, \bar{s}) -\hf P^{\gd \g}I_{\g \ad} (X) I_{\a \gd} (\bar{X}) G^{\ad \a}(X^I;\, v, \bar{v}, s, \bar{s})~.
\eea
The conservation law of $L$ demands that
\bsubeq
\bea
\cQ^2 \widetilde{H} (X, P;\, u, \bar{u}, v, \bar{v})&=& 0~, \label{JJL-z3-1}\\
\bar{\cQ}^2 \widetilde{H} (X, P;\, u, \bar{u}, v, \bar{v})&=& 0~.\label{JJL-z3-2}
\eea
\esubeq
Now, a useful observation is that $\cQ_{\a} X_{\b \bd} = \bar{\cQ}_{\ad} \bar{X}_{\b \bd}= 0$. Hence, in computing \eqref{JJL-z3-1}, it is easier to first express the right-hand side of \eqref{JJL-tildeH-I} in terms of $X$ and $P$ (that is, we replace $\bar{X} = X+ \ri P$). On the other hand, to impose \eqref{JJL-z3-2}, we should first replace $X = \bar{X} - \ri P$ in \eqref{JJL-tildeH-I}. Carrying out this procedure leads to 
\bsubeq
\bea
\widetilde{H}(X, P;\, u, \bar{u}, v, \bar{v}) = \widetilde{F}(X;\, u, \bar{u}, v, \bar{v}) -\hf P^{\ad \a} \widetilde{G}_{\a \ad}(X;\, u, \bar{u}, v, \bar{v})~,
\eea
where
\bea
&&\widetilde{F}(X;\, u, \bar{u}, v, \bar{v}) = \frac{(-1)^{r_2}}{r_2!r_2!} \frac{1}{X^{2(r_2-r_1)}} \big( {u} {X} \pa_{(\bar{s})} \big)^{r_2} \big(\bar{u} {X} \pa_{(s)} \big)^{r_2} F(X;\, v, \bar{v}, s, \bar{s})~,~~~~~\\
&&\widetilde{G}_{\a \ad}({X};\, u, \bar{u}, v, \bar{v}) = \frac{(-1)^{r_2}}{r_2! r_2!} \frac{1}{X^{2(r_2-r_1)}} \big(u {X} \pa_{(\bar{s})}\big)^{r_2} \times \non\\
&&\,\, \times~ \bigg\{ 2 \ri(r_1+1)  \big(\bar{u} {X} \pa_{({s})}\big)^{r_2} \frac{{X}_{\a \ad}}{X^2} F({X}; v, \bar{v}, s, \bar{s}) - 2\ri\, r_2 \big(\bar{u} {X} \pa_{({s})}\big)^{r_2-1} \bar{u}_{\ad} \frac{\pa}{\pa s^{\a}}F({X}; v, \bar{v}, s, \bar{s}) \non\\
&& \qquad +\, \ri \big(\bar{u} X \pa_{({s})}\big)^{r_2} \pa_{\a \ad} F({X}; v, \bar{v}, s, \bar{s}) - \big(\bar{u} X \pa_{({s})}\big)^{r_2} \frac{X_{\a\bd} X_{\b \ad}}{X^2} 
G{}^{\bd \b}(X; v, \bar{v}, s, \bar{s})
\bigg\}~.~~~~~~~~~~~
\eea
\esubeq
In deriving the above expression, the terms proportional to $P^2$ identically vanish due to the conservation conditions on the first two points. Hence, now it is simple to see that \eqref{JJL-z3-1} holds, since $Q_{\a} X_{\b \bd}= 0$ and $\cQ^2 P_{\a \ad} = 0$. In a similar manner, by expanding $X = \bar{X}- \ri P$, we obtain
\bsubeq
\bea
\widetilde{H}(\bar{X}, P;\, u, \bar{u}, v, \bar{v}) = \widetilde{F}(\bar{X};\, u, \bar{u}, v, \bar{v}) -\hf P^{\ad \a} \widetilde{G}_{\a \ad}(\bar{X};\, u, \bar{u}, v, \bar{v})~,
\eea
where
\bea
&&\widetilde{F}(\bar{X};\, u, \bar{u}, v, \bar{v}) = \frac{(-1)^{r_2}}{r_2!r_2!} \frac{1}{\bar{X}^{2(r_2-r_1)}} \big( {u} \bar{X} \pa_{(\bar{s})} \big)^{r_2} \big(\bar{u} \bar{X} \pa_{(s)} \big)^{r_2} F(\bar{X};\, v, \bar{v}, s, \bar{s})~,\label{tF-barQ}\\
&&\widetilde{G}_{\a \ad}(\bar{X};\, u, \bar{u}, v, \bar{v}) = \frac{(-1)^{r_2}}{r_2! r_2!} \frac{1}{\bar{X}^{2(r_2-r_1)}} \big(u \bar{X} \pa_{(\bar{s})}\big)^{r_2} \times \non\\
&&\qquad \times~ \bigg\{ 2 \ri(r_2+1)  \big(\bar{u} \bar{X} \pa_{({s})}\big)^{r_2} \frac{\bar{X}_{\a \ad}}{\bar{X}^2} F(\bar{X}; v, \bar{v}, s, \bar{s}) + 2\ri\, r_2 \big(\bar{u} \bar{X} \pa_{({s})}\big)^{r_2-1} {u}_{\a} \frac{\pa}{\pa \bar{s}^{\ad}}F(\bar{X}; v, \bar{v}, s, \bar{s}) \non\\
&& \qquad \qquad - \big(\bar{u} \bar{X} \pa_{({s})}\big)^{r_2} \frac{\bar{X}_{\a\bd} \bar{X}_{\b \ad}}{\bar{X}^2} 
G{}^{\bd \b}(\bar{X}; v, \bar{v}, s, \bar{s})
\bigg\}~.~~~~~~~~~~~ \label{tG-barQ}
\eea
\esubeq
Without explicitly substituting the expressions for $F$ and $G_{\a \ad}$ in the right-hand sides of \eqref{tF-barQ} and \eqref{tG-barQ}, one can immediately see that $\bar{\cQ}^2 \widetilde{H}(\bar{X}, P;\, u, \bar{u}, v, \bar{v}) = 0$, for $\bar{\cQ}_{\a \ad} \bar{X}_{\b \bd} = 0$ and $\bar{\cQ}^2 P_{\a \ad} = 0$. As a result, conservation on $z_3$ automatically holds for the choice of coefficients \eqref{JJL-parameters}.

The three-point function $\la J_{\a(r_1) \ad(r_1)}(z_1)  J_{\b(r_2) \bd(r_2)}(z_2) L(z_3) \ra$, with $r_1 < r_2$ is thus fixed up to two real parameters: $a_0$ and $b_0$.

\subsubsection{Analysis for $r_1 = r_2$}
In the case when $r_1 = r_2$, the correlator is subject to an extra constraint due to the $1 \leftrightarrow 2$ symmetry. In terms of $F$ and $G_{\a \ad}$, this symmetry is equivalent to imposing
\bsubeq \label{JJL-12symm}
\bea
F(X; \, u, \bar{u}, v, \bar{v}) &=& F(-X;  \,v, \bar{v}, u, \bar{u})~, \\
G_{\a \ad}(X; \, u, \bar{u}, v, \bar{v}) &=& G_{\a \ad}(-X;  \,v, \bar{v}, u, \bar{u})+\,\ri \pa_{\a \ad} F(-X;\, v, \bar{v}, u, \bar{u})~.~~~~~~~
\eea
\esubeq
Both conditions imply that $b_k = 0$, for $k=0,1,\dots, r_1 $. As a result, the correlator 
\bea 
\la J_{\a(r) \ad(r)}(z_1)  J_{\b(r) \bd(r)}(z_2) L(z_3) \ra 
&=&
\frac{1}{(x_{\bar{1} 3} x_{\bar{3} 1} x_{\bar{2} 3} x_{\bar{3} 2})^{r+2}}  \cI_{\a(r) \gd(r)} (x_{1 \bar{3}})\, \cI_{\a(r) \gd(r)} (x_{3 \bar{1}}) \non\\
&&\quad \times~ \cI_{\b(r) \dot{\d}(r)} (x_{2 \bar{3}})\, \cI_{\d(r) \bd(r)} (x_{3 \bar{2}}) \non\\
&&\quad \times~ H^{\gd(r) \g(r),\,\, \dot{\d}(r)\d(r)} (\fn3)~~~~~~~~~
\eea
is determined up to a single real parameter, $a_0$. The explicit solution is given by
\bsubeq
\bea
H(X,P; \, u, \bar{u}, v, \bar{v})
&=& F (X;\, u, \bar{u}, v, \bar{v}) -\hf P^{\ad \a} G_{\a \ad} (X;\, u, \bar{u}, v, \bar{v})~,~~~~~~
\eea
with 
\bea
F(X;\, u, \bar{u}, v, \bar{v}) &=& \frac{1}{X^{2r+2}} \sum_{k=0}^{r} a_k \,(U \hat{X})^{r-k} (V \hat{X})^{r-k} (UV)^{k}~,\\
G_{\a \ad}(X;\, u, \bar{u}, v, \bar{v}) &=& \frac{1}{X^{2r+3}} \bigg[
\hat{X}_{\a \ad}\sum_{k=0}^{r} B_k \,(U \hat{X})^{r-k} (V \hat{X})^{r-k} (UV)^{k}  \non\\
&& +~ u_{\a} \bar{u}_{\ad} \sum_{k=0}^{r-1} C_k \,(U \hat{X})^{r-(k+1)} (V \hat{X})^{r-k} (UV)^{k}  \non\\
&&+~ v_{\a} \bar{v}_{\ad} \sum_{k=0}^{r-1} D_k \,(U \hat{X})^{r-k} (V \hat{X})^{r-(k+1)} (UV)^{k}
\bigg]~.
\eea
\esubeq
Here the parameters are related by
\bsubeq
\bea
B_k &=& -\ri \,(2r+1-k) \,a_k~,\\
C_k &=& D_k = \frac{\ri}{2} (r-k) \,a_k~,
\eea
where, for $k=1,2, \dots r$, we have the recursive relation
\bea
a_k &=& -\frac{(r+1-k)^2}{2k(2r+1-k)}\, a_{k-1}~.
\eea
\esubeq


\section*{Acknowledgements}
We are grateful to Sergei Kuzenko for collaborations at early stages of the work, for valuable discussions, and for comments on the manuscript. 
We would also like to thank Benjamin Stone for useful comments on the main results. The work of E.I.B. is supported in part by the Australian Research Council (ARC), project No. DP200101944. The work of J.H. and G.T.-M. is supported in part by the ARC Future Fellowship FT180100353, and by the Capacity Building Package of the University of Queensland. J.H. and G.T.-M. thank the MATRIX Institute in Creswick for hospitality and support during part of this project.


\begin{footnotesize}

\end{footnotesize}
\end{document}